\documentclass[fleqn]{elsarticle}

\usepackage{amsmath, amssymb, bm} 
\usepackage[utf8]{inputenc} 
\usepackage{float}
\usepackage{indentfirst} 
\usepackage{xcolor} 
\usepackage{setspace} 
\usepackage{booktabs, array} 
\usepackage{multirow} 
\usepackage{mathtools}
\usepackage{lineno,hyperref}
\usepackage{amsmath}
\usepackage{xcolor}
\usepackage[a4paper, left=1in, right=1in, top=1in, bottom=1in]{geometry}

\journal{arXiv}

\makeatletter
\providecommand{\@corref}[1]{} 
\providecommand{\cnotenum}[1]{} 
\makeatother

\begin{document}

\begin{frontmatter}

\title{Topology Optimization for Multi-Axis Additive Manufacturing Considering Overhang and Anisotropy}

\author[unist]{Seungheon~Shin}
\author[unist]{Byeonghyeon~Goh}
\author[unist]{Youngtaek~Oh}
\author[unist]{Hayoung~Chung\corref{mycorrespondingauthor}}
\cortext[mycorrespondingauthor]{Corresponding author}
\ead{hychung@unist.ac.kr}

\address[unist]{Department of Mechanical Engineering, Ulsan National Institute of Science and Technology, 50 UNIST-gil, Ulju-gun, Ulsan, 44919, Republic of Korea}

\begin{abstract}
Topology optimization produces designs with intricate geometries and complex topologies that require advanced manufacturing techniques such as additive manufacturing (AM). However, insufficient consideration of manufacturability during the optimization process often results in design modifications that compromise the optimality of the design. While multi-axis AM enhances manufacturability by enabling flexible material deposition in multiple orientations, challenges remain in addressing overhang structures, potential collisions, and material anisotropy caused by varying build orientations. To overcome these limitations, this study proposes a novel space-time topology optimization framework for multi-axis AM. The framework employs a pseudo-time field as a design variable to represent the fabrication sequence, simultaneously optimizing the density distribution and build orientations. This approach ensures that the overhang angles remain within manufacturable limits while also mitigating collisions. Moreover, by incorporating material anisotropy induced by diverse build orientations into the design process, the framework can take the scan path-dependent structural behaviors into account during the design optimization. Numerical examples demonstrate that the proposed framework effectively derives feasible and optimal designs that account for the manufacturing characteristics of multi-axis AM.
\end{abstract}


\end{frontmatter}

\section{Introduction}
\label{sec:Intro}
Topology optimization is a computational design method that optimizes material distribution within a given design domain to achieve structurally efficient configurations \cite{Bendsoe, Sigmund2013}. This approach often results in intricate geometries and complex topologies, such as Michell structures, which exhibit optimal material distribution under given conditions. However, realizing these intricate structures poses significant challenges for conventional manufacturing methods, necessitating the adoption of advanced techniques capable of handling such complexity \cite{Brackett2011}. Additive manufacturing (AM) offers unparalleled design freedom, enabling the fabrication of intricate geometries that were previously unachievable \cite{Frazier2014,Herzog2016,Ngo2018,Wong2012}. This capability makes AM particularly well-suited for fabricating the complex geometries that arise from topology optimization.

However, while AM facilitates the fabrication of complex structures, its implementation presents significant technical challenges if manufacturing constraints, such as structural manufacturability and process-induced effects, are not explicitly considered \cite{Bayat2023,Ibhadode2023,Zhu2021}. In practice, these challenges are often addressed through preprocessing techniques, such as modifying designs to reduce unsupported regions, or post-processing steps, such as removing auxiliary support structures. However, these methods are not fundamentally suited to ensuring manufacturability and often lead to increased costs, longer production times, and compromises in the performance of the optimized design \cite{Liu2018}. Therefore, rather than relying on such corrective measures, it is essential to develop strategies that integrate manufacturability considerations directly into the design process.
\begin{figure*}[hbt!]
	\centering
	\includegraphics[scale=0.9]{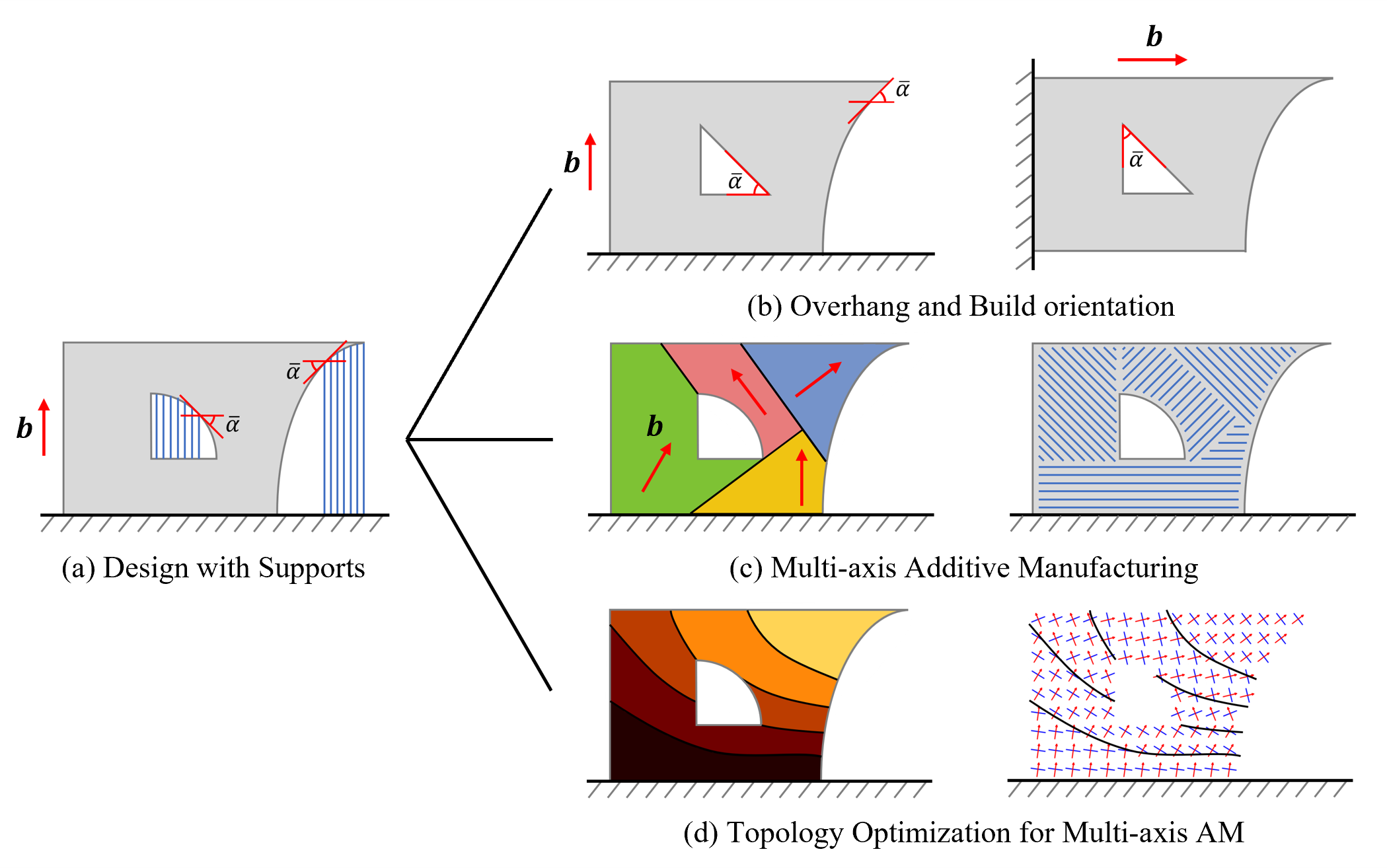}
	\caption{Design approaches for AM: (a) design with supports, where $\bar{\alpha}$ represents the overhang angle threshold and the red arrows $\bm{b}$ denote the build orientation, (b) designs considering overhang angles and build orientations, (c) designs for multi-axis AM with divided build orientations and paths, and (d) proposed space-time topology optimization for multi-axis AM, where red arrows indicate the build orientation and the blue lines represent the material orientation.}
	\label{fig.1}
\end{figure*}

In response to this need, numerous studies have focused on improving manufacturability while also accounting for process-induced effects arising from the unique characteristics of AM processes. As highlighted in recent reviews \cite{Bayat2023,Ibhadode2023,Liu2018,Zhu2021}, issues such as length scale, overhang structures, and anisotropic material properties often complicate the direct realization of optimized designs. Among these challenges, overhang structures have received particular attention due to their critical impact on manufacturability and structural integrity. As illustrated in Fig. \ref{fig.1}(a), these structures often necessitate additional supports during fabrication, leading to increased material usage and production costs \cite{Jiang2018}. To address this, research efforts have proposed incorporating overhang angle constraint directly into topology optimization frameworks as depicted in Fig. \ref{fig.1}(b) \cite{Allaire2017,Gaynor2016,Langelaar2017,Qian2017,Wang2020a,Zhang2019,Zou2021}. Nevertheless, these methods typically assume a single-build orientation, restricting design freedom and thus limiting their applicability to complex geometries. In a similar manner, the anisotropic material properties inherent in metal AM pose challenges to achieving optimal mechanical performance. Recent topology optimization methods have started addressing these challenges by incorporating anisotropy considerations into the design process \cite{Bruggi2021,Mishra2022,Wu2024,Zhou2021}. However, these approaches remain insufficient for fully ensuring manufacturability, particularly when addressing complex interactions between material anisotropy and other manufacturing constraints in AM.

Recent developments in design for AM have explored innovative approaches to address manufacturing constraints, expanding the potential of AM-enabled structures. For example, multi-axis additive manufacturing enables fabrication from multiple orientations, minimizing the need for support structures and improving geometric adaptability \cite{Jiang2021,Tang2024}. Leveraging this capability, design strategies utilize post-processing to divide an existing configuration into smaller, manufacturable components, enabling sequential fabrication as shown in Fig. \ref{fig.1}(c) \cite{Dai2018,Gao2019,Wu2020,Xu2019}. Similarly, the space-time topology optimization method integrates spatial and temporal parameters into the optimization process, incorporating the manufacturing sequence in multi-axis AM during design process \cite{Wang2020b,Wang2024,Wu2024}. While these methods represent progress in addressing AM constraints, they primarily focus on component segmentation and process sequencing rather than directly embedding manufacturability considerations, such as overhang angle constraint, into the topology optimization formulation.

To overcome the limitations of existing approaches, this study proposes a novel topology optimization framework that incorporates space-time topology optimization techniques to address key manufacturing constraints in AM. The proposed method explicitly considers the sequencing of the AM process by dynamically adjusting the build orientation and the amount of deposited material in discrete timesteps during the design phase as illustrated in Fig. \ref{fig.1}(d). This enables the generation of self-supporting structures that satisfy overhang angle constraint without requiring additional support structures. Furthermore, the framework accounts for the anisotropic material properties inherent to metal additive manufacturing, optimizing the mechanical performance of the structure under realistic operating conditions. By integrating these considerations directly into the optimization process, the method provides a comprehensive approach to achieving manufacturable and efficient designs.

This manuscript is organized as follows. Section 2 presents the mathematical formulation of the proposed topology optimization framework, including the incorporation of manufacturing constraints and the modeling of material anisotropy. Section 3 details the implementation of the framework and demonstrates its effectiveness. Section 4 discusses the results, highlighting the advantages and limitations of the proposed method, and concludes with potential future research directions.
\section{Method} \label{sec:Method}
\begin{figure*}[bt!]
	\centering
	\includegraphics[scale=0.9]{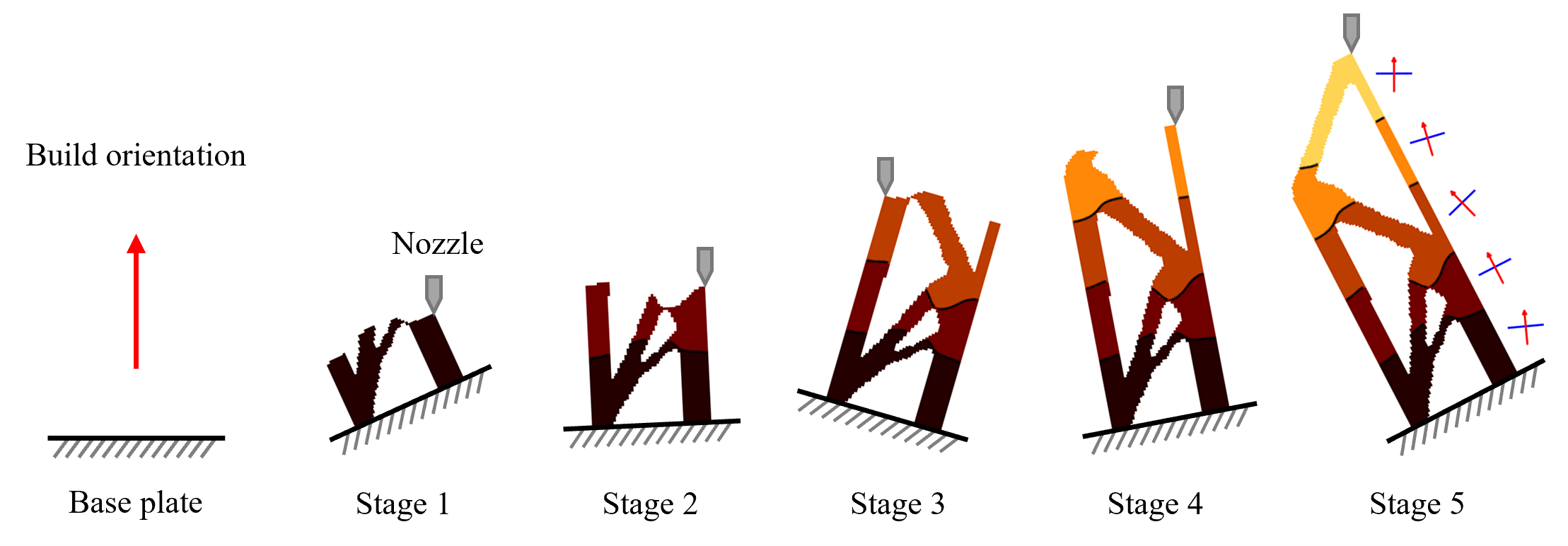}
	\caption{The concept of space-time topology optimization for multi-axis AM. Each building stage has a different platform orientation, affecting the material's principal axis and overhang behavior. The red arrows denote the build orientation, while the blue lines indicate the material orientation at each stage.}
	\label{fig.2}
\end{figure*}

In this section, the proposed topology optimization method for multi-axis AM is introduced. This study adapts space-time topology optimization \cite{Wang2020b} to define intermediate structures formed at different printing stages, as illustrated in Fig. \ref{fig.2}, which shows stage-wise material depositions and changes in build orientation. 
Using this method, the spatial distribution of material density $\bm{\rho}$ and the stage-wise build orientation $\bm{\theta}$ are determined, leading to the intermediate structures shown in Fig. \ref{fig.2}, which parameterize the stage-wise manufacturing process. Moreover, the amount of materials deposited at each stage is also determined. All intermediate designs not only satisfy manufacturing constraints such as material overhang but also account for process-induced properties like material anisotropy. These considerations ensure that the intermediate structures satisfy multi-axis AM constraints while achieving structural optimality. 

The following discussion focuses on the fundamental principles and applications of the proposed method, including space-time topology optimization \cite{Wang2020b}, the overhang angle, and material anisotropy, emphasizing its relationship with build orientations and the pseudo-time field.
\subsection{Design parameterization of intermediate structures}
\label{sec:space-time_topopt}
The following section briefly introduces the space-time topology optimization method proposed by Wang et al. \cite{Wang2020b}, which is adapted herein to parametrize intermediate designs.
The design variables are the material density field $\bm{\psi}$ and the pseudo-time field $\bm{\tau}$, which are piecewise constant variables based on a finite element discretization of the design space. 
Both design variables range from 0 to 1, which is typical in density-based topology optimization. 
The density variable $\rho_e$ indicates whether a single element $e$ is filled with material (i.e., $\rho_e=1$) or void (i.e., $\rho_e=0$), and its distribution determines the overall topology of the resulting structure. Continuous values of $\tau_e$ indicate the printing sequence, where $\tau_e = 0$ represents the initial stage (i.e., $j=0$) and $\tau_e = 1$ represents the final stage (i.e., $j=N$). Notably, unlike $\rho_e$, the pseudo-time variable $\tau_e$ is not strictly binary.

The intermediate structure at stage $j$ is parameterized using the set of design variables via the element-wise density variable $\rho_e^{\{j\}}$, which represents the density of material deposited at or before specific stage $j$ at element $e$:
\begin{equation}
    \rho_e^{\{j\}} = \rho_e(\psi_e)\cdot\bar{t}_e^{\{j\}}(\tau_e), 
\end{equation}
where $\rho_e(\psi_e)=\rho_{\text{min}} + (1 - \rho_{\text{min}})\cdot \bar{\psi}_e(\psi_e)$ represents the time-invariant projected density with a small fictitious value $\rho_{\text{min}}=10^{-3}$ to ensure numerical stability, and $\bar{t}_e^{\{j\}}$ denotes the truncated pseudo-time, which equals unity only when the material is deposited at stage $j$ or earlier. Ideally, both variables that parameterize the stage-wise structure (i.e., $\rho_e$ and $\bar{t}_e^{\{j\}}$) are binary.
To enforce this binary nature, the Heaviside projection is applied:
\begin{equation}
\begin{aligned}
    \bar{\psi}_e = \frac{\tanh(\beta_d \eta) - \tanh(\beta_d (\tilde{\psi}_e - \eta))}{\tanh(\beta_d \eta) - \tanh(\beta_d (1 - \eta))}, 
    \quad   
    \bar{t}_e^{\{j\}} = 1 - \frac{\tanh(\beta_t \tau_j) - \tanh(\beta_t (\tilde{\tau}_e - \tau_j))}{\tanh(\beta_t \tau_j) - \tanh(\beta_t (1 - \tau_j))},
\end{aligned}
\label{eqn:projection}
\end{equation}
where the soft Heaviside functions, controlled by the sharpness parameters $\beta_d$ and $\beta_t$, are used to obtain continuous and differentiable variables. For these functions, the threshold value of the density field $\eta$ is set to 0.5, while that of the time field $\tau_j$ is defined as $j/N$, where $N$ denotes the number of stages, assuming a uniform interval between truncated times. However, it should be noted that the uniformity of the time interval does not imply a uniform amount of material deposition at each stage, as will be shown in the later sections.

In the present work, we adopt a typical regularization filter with a radius $r_d$ and linearly decaying weight $w$ to obtain continuous and smooth fields for density and time (i.e., $\tilde{\psi}_e$ and $\tilde{\tau}_e=t_e$). Here, $\tilde{\Xi}$ denotes the filtered variable of $\Xi$, which is defined as follows:
\begin{equation}
    \tilde{\Xi}_e = \frac{\sum_{i \in \mathcal{S}_e} w(\mathbf{x}_i, r_d) \Xi_i}{\sum_{i \in \mathcal{S}_e} w(\mathbf{x}_i, r_d)},
    \quad
    w(\mathbf{x}_i, r_d) = r_d - \|\mathbf{x}_i - \mathbf{x}_e\|,
\label{eqn:filtering}
\end{equation}
where $\mathbf{x}_e$ is the centroid of element $e$, and $\mathbf{x}_i$ is the centroid of a neighboring element within the set \( \mathcal{S}_e = \{i \mid w(\mathbf{x}_i, r) > 0\} \). 

As a result, the optimized pseudo-time field divides the density into $N$ stages, representing the fabrication sequence in AM. This method is essential for reflecting the characteristics of multi-axis AM, namely non-constant build orientations that determine manufacturability and material properties.
\subsection{Overhang angle}
During the multi-axis AM process, all intermediate structures should be manufactured such that the angle formed between each of their build orientations and the boundary normal of the structure is greater than a certain threshold, denoted as $\bar{\alpha}$. Incorporating such a manufacturing constraint into the design process is challenging, as it requires not only identifying the printing front but also determining the normal of intermediate structures, which are pixelated and therefore not continuous.

To address this challenge, the overhang angle is calculated based on the boundary normal as $\alpha = cos^{-1}({\nabla \rho_e}/{\|\nabla \rho_e\|_2} \cdot \bm{b})$, where the build orientation is defined as $\bm{b}=(cos\theta, sin\theta)$. Here, $\theta$ denotes the angle between the build orientation $\bm{b}$ and the normal vector of the base plate. The computed density gradients are then used to determine the gradient of each element and to formulate the overhang angle constraint as follows.
\begin{equation}
\begin{aligned}
    \alpha \geq \bar{\alpha}
\end{aligned}
\label{eq:overhang}
\end{equation}
Here, $\alpha$ represents the overhang angle, and $\bar{\alpha}$ denotes the threshold for the overhang angle. This constraint must be satisfied by every element within the density domain. 

\begin{figure*}[hbt!]
	\centering
	\includegraphics[scale=0.9]{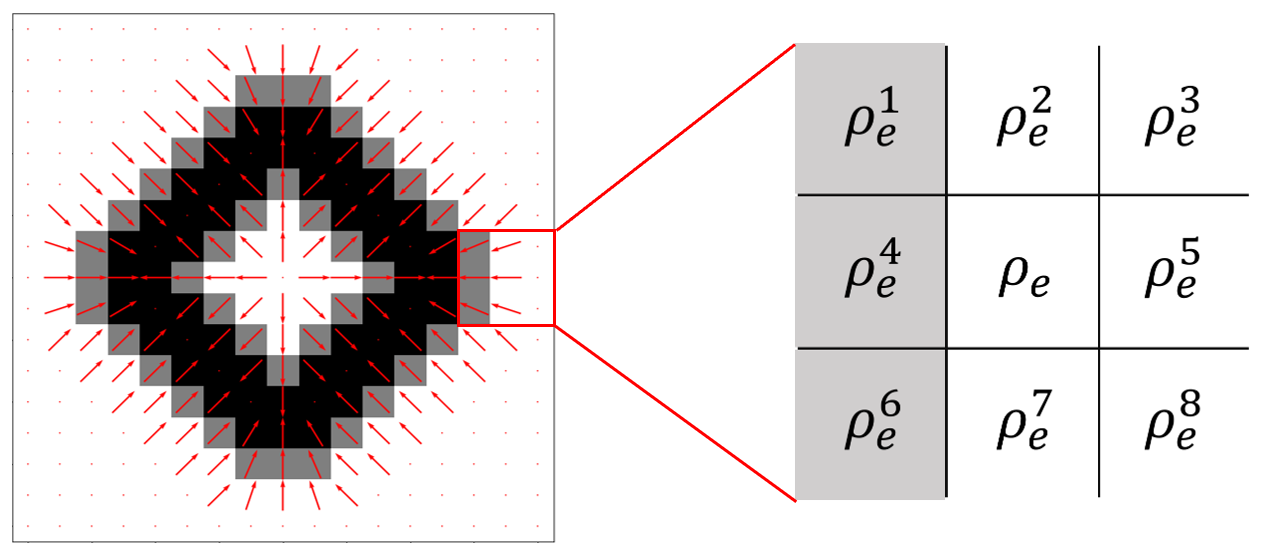}
	\caption{Calculated density gradients using the Sobel operator. The figure shows gradients with fixed magnitudes and the corresponding element-wise numbering for the highlighted region.}
	\label{fig.3}
\end{figure*}

In this study, the density gradient method, which is a well-known image processing technique used to identify both the boundary of a pixelated image and its normal, is employed to address these challenges. The Sobel operator \cite{Ansari2017, Vincent2009} is used to compute the field of density gradient $\nabla\rho_e$ for each element at the specific build stage:
\begin{equation}
\begin{aligned}
    \nabla \rho_e = G_i\bm{e}_i, 
    \label{eq:overhang 1}
\end{aligned}
\end{equation}
where $G_i$ is the gradient in the direction $i$, calculated as the sum of all elements of $\bm{C}_i$, which reads:
\begin{equation}
\begin{aligned}
    \bm{C}_i = \bm{h}_i\odot \bm{A}_e, 
    \label{eq:overhang 2}
\end{aligned}
\end{equation}
where the operator $\odot$ denotes element-wise multiplication (Hadamard product). 
The term $h_i$ represents the Sobel kernel, and $A_e$ represents the density field as follows:
\begin{equation}
\begin{aligned}
    \bm{h}_1 &= 
    \begin{bmatrix}
    -1 & 0 & 1 \\
    -2 & 0 & 2 \\
    -1 & 0 & 1
    \end{bmatrix}, \quad
    \bm{h}_2 = 
    \begin{bmatrix}
    1 & 2 & 1 \\
    0 & 0 & 0 \\
    -1 & -2 & -1
    \end{bmatrix}, \quad
    \bm{A}_e = 
    \begin{bmatrix}
    \rho_e^1 & \rho_e^2 & \rho_e^3 \\
    \rho_e^4 & \rho_e   & \rho_e^5 \\
    \rho_e^6 & \rho_e^7 & \rho_e^8
    \end{bmatrix},
    \label{eq:overhang 3}
\end{aligned}
\end{equation}
The term $\rho_e^n$ represents the density of the nearest neighboring elements of element $\rho_e$, as shown in Fig. \ref{fig.3}. 

The density gradients obtained from Eqs. (\ref{eq:overhang 1})--(\ref{eq:overhang 3}) are illustrated in Fig. \ref{fig.3}. The $\nabla \rho_e$ vectors, shown as red arrows within the corresponding elements, represent the normals. The boundary is identified in regions where the density is non-uniform. In regions with similar density values, the gradient approaches zero (i.e.,  $\nabla \rho_e \approx (0, 0)$). This confirms that the Sobel operator effectively detects density transitions, ensuring that overhang angles are calculated only where necessary.
\subsection{Material anisotropy}
Material anisotropy is commonly observed in additively manufactured structures \cite{Bruggi2021,Hadjipantelis2022,Kyvelou2020} as the primary material axis is determined by deposition orientation. In this study, the time field and build orientation are utilized to determine the material orientation for each stage and to integrate it into the optimization process. 
\begin{figure*}[bt!]
	\centering
	\includegraphics[scale=0.9]{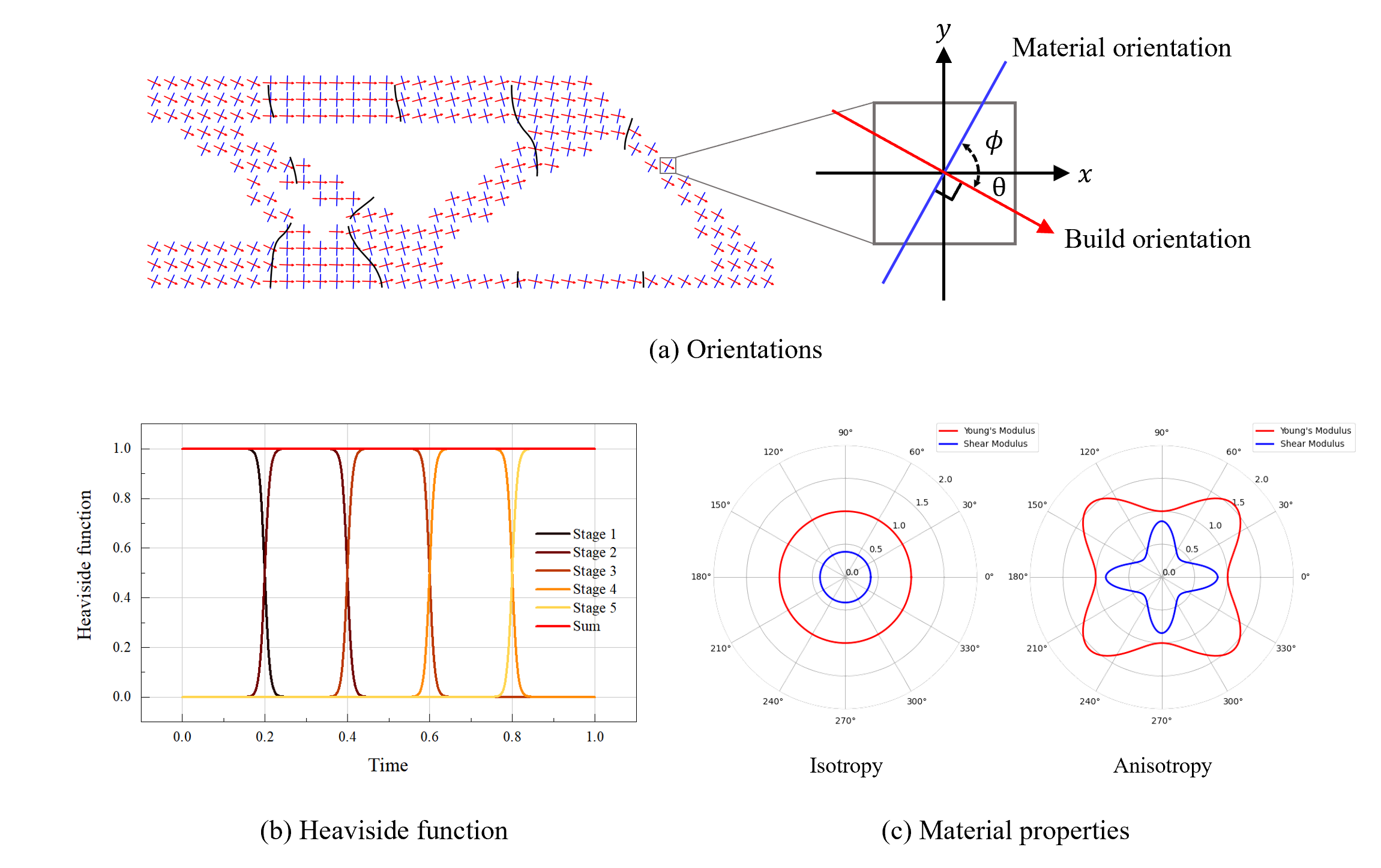}
	\caption{(a) Build and material orientations in multi-axis AM, (b) Heaviside function defined over the time field for $N=5$, and (c) Young's modulus and shear modulus with respect to orientation angle for isotropic and anisotropic materials.}
	\label{fig.4}
\end{figure*}
For simplicity, in the current multi-axis AM topology optimization framework, the material orientation is aligned perpendicular to the build orientation at each stage (Fig. \ref{fig.4}(a)) \cite{Guo2025,Ye2023}. A differentiable form of the element-wise material orientation $\phi_e$ is calculated at each stage:
\begin{equation}
\begin{aligned}
    \phi_e = \sum_{j=1}^N (\bar{t}_e^{\{j\}} - \bar{t}_e^{\{j-1\}})(\theta_j + \frac{\pi}{2}),
    \label{eq:material orientation}
\end{aligned}
\end{equation}
where $\bar{t}_e^{\{j\}} - \bar{t}_e^{\{j-1\}}$ approximates the indicator function for each stage (Fig. \ref{fig.4}(b)). Here, $\pi/2$ is added because the material orientation is assumed to be perpendicular to the build orientation. Additionally, using summation instead of the indicator function prevents discontinuities in material orientation at the Heaviside function threshold, thereby ensuring differentiability.

The element-wise constant material orientation $\phi_e$, calculated using this formulation, is incorporated into the stress-strain matrix $D$ through Hooke’s law for orthotropic materials. This ensures that the anisotropic properties of the material, influenced by the build orientation at each stage, are appropriately reflected in the structural analysis and optimization process. The stress-strain matrix $\mathbf{D}_e(\phi_e)$ is defined as follows:
\begin{equation}
\begin{aligned}
    \mathbf{D}_e(\phi_e) = \mathbf{T}(\phi_e) \mathbf{D}_0 \mathbf{T}(\phi_e)^T
\label{eq:TDT}
\end{aligned}
\end{equation}

For simplicity, the material properties are approximated as cubic orthotropic material properties in this study, following the approach in \cite{Wu2024}. Considering the plane stress condition, the stress-strain matrix $\mathbf{D}_0$ is defined as follows:
\begin{equation}
\begin{aligned}
    \mathbf{D}_0 = \frac{1}{1 - \nu_{12} \nu_{21}}
    \begin{bmatrix}
    E_{11} & \nu_{21} E_{11} & 0 \\
    \nu_{12} E_{22} & E_{22} & 0 \\
    0 & 0 & G_{12}(1 - \nu_{12} \nu_{21})
    \end{bmatrix}
\end{aligned}
\end{equation}
The terms $E$, $G$ and $\nu$ represent Young’s modulus, shear modulus, and Poisson’s ratio, respectively. The transformation matrix is defined as follows:
\begin{equation}
\begin{aligned}
    \mathbf{T}(\phi_e) =
    \begin{bmatrix}
    \cos^2(\phi_e) & \sin^2(\phi_e) & -2 \cos(\phi_e) \sin(\phi_e) \\
    \sin^2(\phi_e) & \cos^2(\phi_e) & 2 \cos(\phi_e) \sin(\phi_e) \\
    \cos(\phi_e) \sin(\phi_e) & -\cos(\phi_e) \sin(\phi_e) & \cos^2(\phi_e) - \sin^2(\phi_e)
    \end{bmatrix}
\end{aligned}
\end{equation}
By utilizing the transformation matrix, material anisotropy relative to the build orientation can be incorporated into the design process.
\begin{table}[bt!]
    \centering
        \begin{small}
        \caption{Material properties used in this study.}
        \label{table.1}
        \begin{tabular}{@{}l l l l l l l @{}}
            \toprule
            \multirow{2}{*}
            & \text{Type} & $E_{11}$ & $E_{22}$ & $G_{12}$ &$\nu_{12}$ & $\nu_{21}$\\ 
            \cmidrule(lr){2-7}
            & \text{Isotropic} &$1$ &$1$ &$0.385$ &$0.3$ &$0.3$ \\ 
            & \text{Anisotropic} &$1$ &$1$ &$0.849$ &$0.3$ &$0.3$ \\ 
            \bottomrule
        \end{tabular}
        \end{small}
\end{table}

The material properties used in this study are depicted in Fig. \ref{fig.4}(c) and Table \ref{table.1}, with the anisotropic properties adopted from experiments \cite{Kyvelou2020,Wu2024}. Both isotropic and anisotropic materials are scaled so that Young’s modulus along the principal axis is 1.0 for both material types.
\subsection{Manufacturing constraints}
This section introduces the manufacturing constraints integrated into the topology optimization framework to reflect the manufacturing characteristics of multi-axis AM.
\subsubsection{Continuity constraint}
To prevent manufacturing issues such as isolated material deposition or fabrication within already constructed structures, the continuity of the time field is enforced during optimization through a continuity constraint derived from space-time topology optimization \cite{Wang2020b}. This constraint ensures sequential fabrication by enforcing that each element has at least one neighboring element with a smaller pseudo-time value. The continuity constraint can be expressed as follows:
\begin{equation}
\begin{aligned}
    \frac{1}{n(\mathcal{M})} \sum_{e \in \mathcal{M}} \left\|t_e - \frac{\sum_{i \in \mathcal{N}_e} t_i}{n(\mathcal{N}_e)}\right\|^2 \leq \gamma,
\end{aligned}
\end{equation}
where $n(\bullet)$ represents the number of elements in a given set. The domain set $\mathcal{M}$ includes all active elements except the fixed-time elements in the starting region, while $\mathcal{N}_e$ denotes the set of neighboring elements of element $e$. The parameter $\gamma$ represents the continuity constraint and is assigned a very small value, determined heuristically.
\subsubsection{Overhang angle constraint}
To maintain the overhang angle above a specific value, as shown in Eq. (\ref{eq:overhang}), the overhang angle constraint function, adapted from \cite{Wang2020a}, is incorporated into the multi-axis AM framework and applied to the design domain. Using this formulation, the overhang angle constraint for each stage’s build orientation can be quantified as a single value, while incorporating self-support conditions from previously fabricated structures, as follows:
\begin{equation}
\begin{aligned}
    P_{\bar{\alpha}} = 
    {\sum_{e \in \mathcal{M}} 
              \sum_{j=1}^N \left(
              H\left(\xi_e{(\bm{b}^{\{j\}}, 
              \nabla\rho_e^{\{j\}})}\right) 
              \bm{b}^{\{j\}} \cdot \nabla\rho_e^{\{j\}}
              (\rho_e^{\{j\}} - \rho_e^{\{j-1\}})\right)}
    ={\sum_{e \in \mathcal{M}} 
              \sum_{j=1}^N P_e^{\{j\}}}
    ={\sum_{e \in \mathcal{M}}P_e}
    \leq \bar{P}_{\bar{\alpha}},
\end{aligned}
\label{eq:overhang_constraint}
\end{equation}
where $\bar{P}_{\bar{\alpha}}$ is the overhang angle constraint parameter, a small value determined heuristically. The $H(\xi_e{(\bm{b}^{\{j\}},\nabla\rho_e^{\{j\}})})$ is defined as follows:
\begin{equation}
\begin{aligned}
    H\left(\xi_e(\bm{b}^{\{j\}}, \nabla \rho_e^{\{j\}})\right)= {(1 + e^{-\beta \xi_e})}^{-1},
\end{aligned}
\end{equation}
where, $\xi_e=(\bm{b}^{\{j\}} \cdot {\nabla \rho_e^{\{j\}}}/{\|\nabla \rho_e^{\{j\}}\|_2} - \cos(\bar{\alpha}))$, and $\beta$ controls the slope of the function transition.
\begin{figure*}[hbt!]
	\centering
	\includegraphics[scale=0.9]{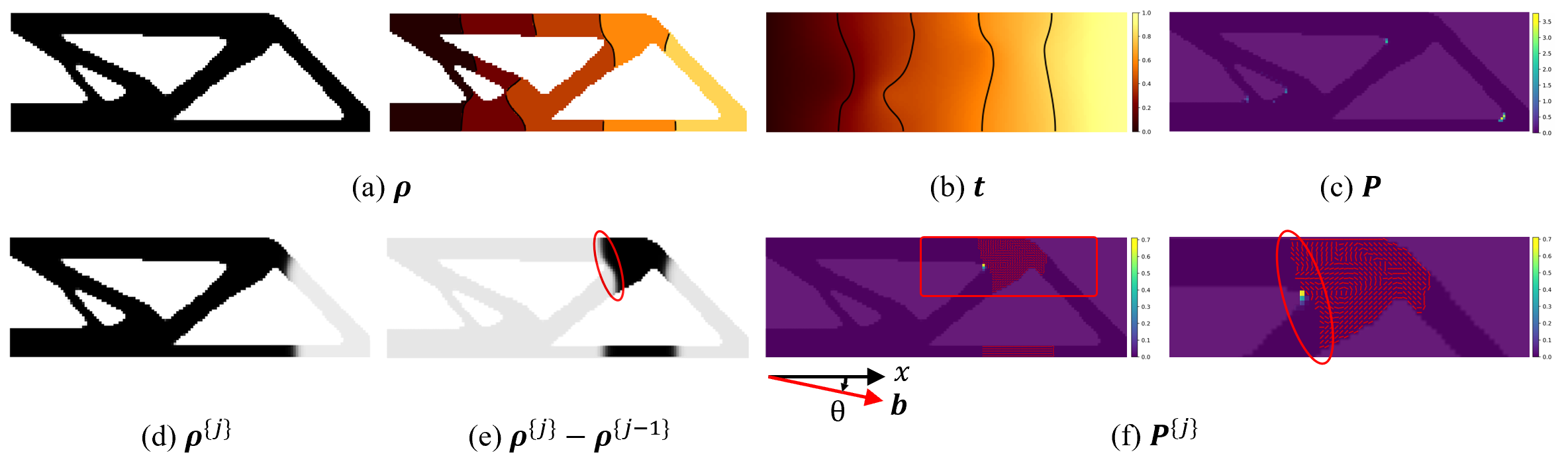}
	\caption{Overhang angle constraint in multi-axis AM: (a) density distribution for the entire domain, (b) time field, (c) overhang angle constraint values for the entire domain, (d) density distribution at the stage $j$, (e) density difference between consecutive stages, and (f) overhang angle constraint values for the stage $j$, where the red lines in each element indicate the density boundaries.}
	\label{fig.5}
\end{figure*}

Figure \ref{fig.5} illustrates an example of the overhang angle constraint application. As shown in Fig. \ref{fig.5}(a), the density distribution is divided into multiple stages according to the time field in Fig. \ref{fig.5}(b). The computed values of $\bm{P}$ for the entire domain are shown in Fig. \ref{fig.5}(c). To account for self-support conditions, it is worth noting that the density gradient is computed using $\rho^{{\{j\}}}$, as shown in Fig. \ref{fig.5}(d), while the density product is based on $\rho^{{\{j\}}} - \rho^{{\{j-1\}}}$, as depicted in Fig. \ref{fig.5}(e). This approach ensures that the constraint is applied only to structures generated in the current stage, while also addressing self-support conditions in previously fabricated regions, highlighted in red in Fig. \ref{fig.5}(e). As a result, no overhang violations occur in regions fabricated in stage $j-1$, as demonstrated by $\bm{P}^{\{j\}}$ in Fig. \ref{fig.5}(f). The unclear boundaries in Fig. \ref{fig.5}(d) and (e), as well as the misalignment between $\bm{P}^{\{j\}}$ and the density boundary within elements in Fig. \ref{fig.5}(f), are due to the non-binary transitions of the Heaviside function. Residual artifacts are present solely for visual clarity.
\section{Topology optimization} \label{sec:Topology optimization}
This section outlines the topology optimization problem formulation and its implementation. Finite element analysis and topology optimization are integrated within the OpenMDAO-TopOpt framework, enabling efficient coupling between analysis and optimization \cite{Chung2019}.
\subsection{Problem formulation}
The objective of the optimization problem is to maximize the stiffness of the final structure for a given amount of allowed material $V_0$, while accounting for overhang angle constraint and material anisotropy. The formulation of the optimization problem is as follows:
\begin{subequations}
\begin{align}
    \min_{\bm{\psi}, \bm{\tau}, \bm{\theta}} \quad & c = \mathbf{U}^T \mathbf{F} \tag{16(a)}\label{eq:objective_function} \\  
    \text{s.t.} \quad 
    & \mathbf{K}(\bm{\rho}) \mathbf{U} = \mathbf{F}, \quad  \text{(Isotropy)} \tag{16(b)}\label{eq:isotropy}\\ 
    & \mathbf{K}(\bm{\rho}, \bm{t}, \bm{\theta}) \mathbf{U} = \mathbf{F}, \quad  \text{(Anisotropy)} \tag{16(c)}\label{eq:anisotropy} \\  
    & \sum_e \rho_e v_e -V_0 \leq 0, \tag{16(d)}\label{eq:volume_constraint} \\  
    & \frac{1}{n(\mathcal{M})} \sum_{e \in \mathcal{M}} 
      \left\|t_e - 
      \frac{\sum_{i \in \mathcal{N}_e} t_i}{n(\mathcal{N}_e)}\right\|^2 -\gamma\leq 0 \tag{16(e)}\label{eq:continuity_constraint}\\   
    & 
    {\sum_{e \in \mathcal{M}} 
              \sum_{j=1}^N \left(
              H\left(\xi_e{(\bm{b}^{\{j\}}, 
              \nabla\rho_e^{\{j\}})}\right) 
              \bm{b}^{\{j\}} \cdot \nabla\rho_e^{\{j\}}
              (\rho_e^{\{j\}} - \rho_e^{\{j-1\}})\right)}
      -\bar{P}_{\bar{\alpha}} 
      \leq 0, \tag{16(f)}\label{eq:overhang_constraint2}\\   
    & 0 \leq \psi_e \leq 1, \tag{16(g)} \label{eq:psi_range}\\  
    & 0 \leq \tau_e \leq 1,  \tag{16(h)}\label{eq:tau_range}\\ 
    & 
      \theta_0 - 
      \frac{\pi}{2} 
      \leq 
      \theta_j 
      \leq 
      \theta_0 + 
      \frac{\pi}{2}  \tag{16(i)}\label{eq:build_orientation_range}
\end{align}
\end{subequations}
Here, $c$ represents the end compliance of the final structure, as formulated in Eq. (\ref{eq:objective_function}). The finite element method is used to analyze the mechanical displacement $\mathbf{U}$ of the optimized structure under a given force $\mathbf{F}$. The stiffness matrix $\mathbf{K}$ is determined by the projected density $\rho_e$ as $\mathbf{K}_e = \mathbf{K}_0 \rho_e^p$, where $\mathbf{K}_0$ represents the stiffness matrix for fully dense material. A penalization parameter $p=3$ was used in this study, which is commonly applied in similar works. When material anisotropy is considered, the build orientation $\bm{\theta}$ also influences $\mathbf{K}_0$, as described in Eq. (\ref{eq:material orientation}) and Eq. (\ref{eq:TDT}). Consequently, isotropy and anisotropy are detailed in Eq. (\ref{eq:isotropy}) and Eq. (\ref{eq:anisotropy}), respectively. 

Unlike conventional space-time topology optimization, which imposes a volume constraint on each intermediate structure \cite{Guo2025, Wang2020b}, the proposed approach applies the volume constraint only to the final structure, as defined in Eq. (\ref{eq:volume_constraint}). This allows the volume of each intermediate structure at different stages to vary freely according to the build orientation, enabling the optimization process to identify the optimal solution. Additionally, to account for the characteristics of multi-axis AM, the time continuity constraint (Eq. (\ref{eq:continuity_constraint})) and the overhang angle constraint (Eq. (\ref{eq:overhang_constraint2})) are incorporated.

The design variables in this framework include the density $\bm{\psi}$, time $\bm{\tau}$, and build orientation $\bm{\theta}$. The bounds for the density and the time design variable are constrained between 0 and 1, as defined in Eq. (\ref{eq:psi_range}) and Eq. (\ref{eq:tau_range}). Likewise, the build orientation is restricted to a range of $\pm \pi / 2$ from the initial orientation, as specified in Eq. (\ref{eq:build_orientation_range}), to account for the rotational limits of multi-axis AM.
\subsection{Implementation detail}
The optimization problem involves multiple constraints, requiring a complex sensitivity analysis based on the adjoint method. To implement this effectively, we adopted OpenMDAO-TopOpt \cite{Chung2019}. This approach leverages a modular architecture, which greatly enhances the reusability and reconfigurability of the encapsulated objects \cite{Gray2019}. These features are particularly useful in the present study as they facilitate efficient transitions between isotropic and anisotropic conditions within the framework, since the total derivatives of the system are automatically calculated using only the partial derivatives of constituting components. Each component is configured to compute partial derivatives with respect to its inputs and outputs, which are then assembled using the MAUD architecture to obtain the total derivatives necessary for the optimization process \cite{Hwang2018}. This automated computation of analytic total derivatives provides significant advantages in handling the complexities of large-scale design problems.
\begin{figure*}[hbt!]
	\centering
	\includegraphics[scale=0.18]{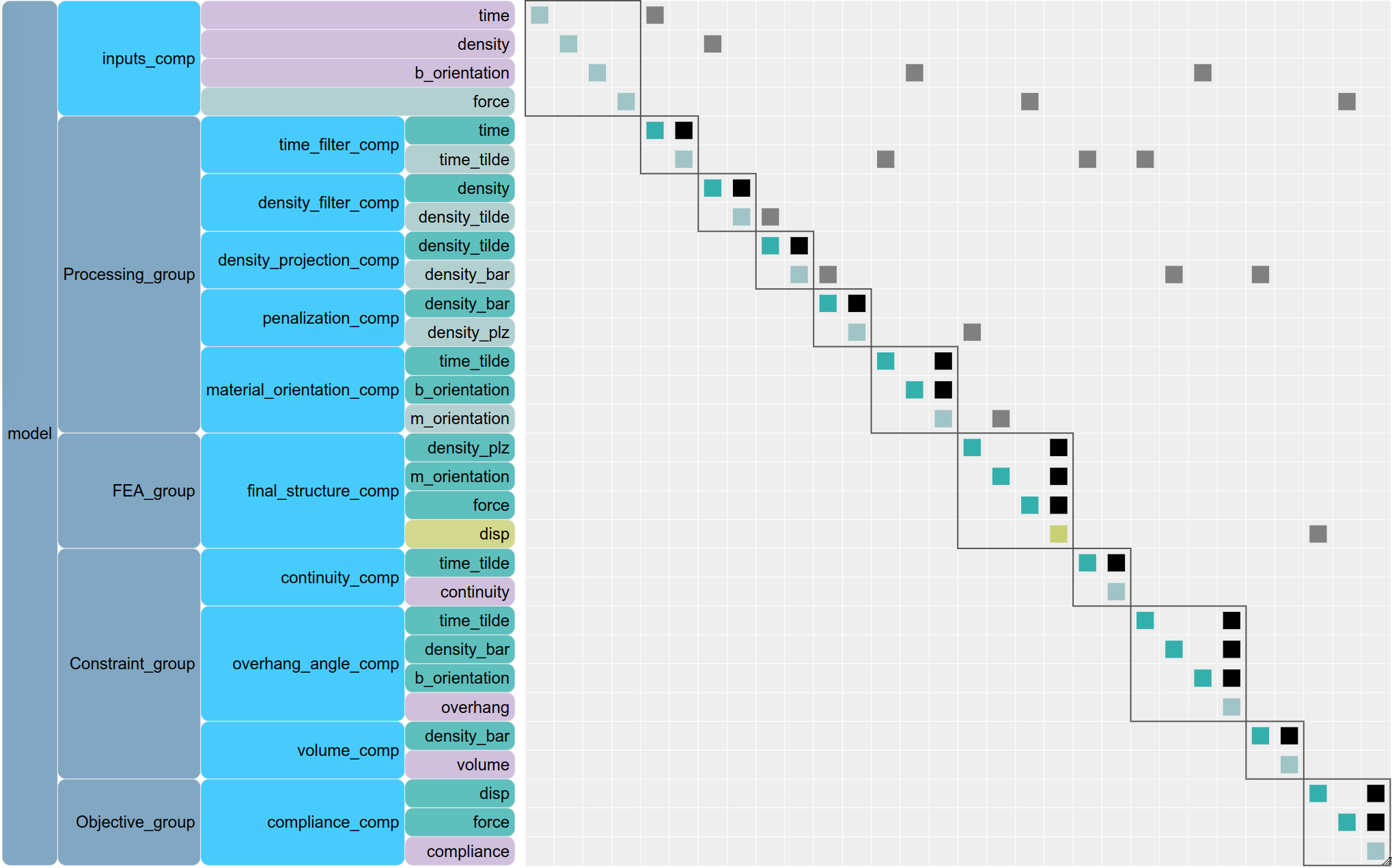}
	\caption{N$^2$ diagram illustrating the architecture of the topology optimization problem, including design variables, constraints, the objective function, and their interconnections.}
	\label{fig.6}
\end{figure*}

The N$^2$ diagram for the current anisotropic framework is presented in Fig. \ref{fig.6}. This diagram, commonly used in multidisciplinary design optimization (MDO), visualizes system components, their connectivity, and input-output variable relationships. The components are grouped based on their roles in the topology optimization process, as follows: design variables (\textit{Input\_group}), filtering and projection of $\bm{\psi}$ and $\bm{\tau}$ (\textit{Processing\_group}), finite element analysis (\textit{FEA\_group}), constraints (\textit{Constraint\_group}), and the objective function (\textit{Objective\_group}). Each group consists of multiple components that serve specific functions. The interactions of components, defined through input-output relationships and partial derivatives, are systematically managed within the framework. For the isotropic framework, the structure remains unchanged except for the exclusion of the material\_orientation\_comp, a component that defines the material orientation for each element. By defining functions and their corresponding partial derivatives for each component, the optimization problem is systematically formulated, effectively managing the interactions among design variables, objective, and constraints while ensuring flexibility and scalability in the process.

For the optimization, SNOPT (Sparse Nonlinear Optimizer) was used. SNOPT is a gradient-based algorithm designed for large-scale nonlinear optimization problems \cite{Gill2005}. This capability makes it well-suited for solving structural topology optimization problems involving a large number of design variables and constraints. To ensure both accuracy and stable convergence, the optimization parameters were set as follows: a Major Optimality Tolerance of $10^{-5}$, a Major Iterations Limit of $500$, and a Major Step Limit of $0.05$.
\section{Numerical examples} \label{sec:Numerical examples}
In this section, several numerical examples are presented to demonstrate the effectiveness of the proposed approach in solving topology optimization problems that consider stage-specific build orientations and manufacturing constraints. The applicability of the proposed method, which designs the spatiotemporal distributions of both material distribution and build orientation, is illustrated in the first example. 
The optimization results are further analyzed to elucidate the effect of key parameters considered in the problem definition, namely the continuity constraint $\gamma$ and the stage-specific or global material volume constraint. Additionally, a design optimization problem considering material anisotropy is presented in the subsequent example, highlighting how the framework accommodates material anisotropy depending on stage-specific build orientations. The section concludes with a model featuring an initial cutout, showcasing the adaptability of the method when the overhang angle constraint is initially violated.

Unless otherwise stated, all numerical examples follow the analysis setup below. The 2D finite element analysis is performed under plane stress conditions using a structured quadrilateral mesh. The filter radius for both density and time is set to five element lengths i.e., $r_d=5.0$. The Heaviside projection parameters, $\beta_d$ and $\beta_t$, are initialized at 20 for both density and time and are incremented by three every five iterations starting from the 40th iteration including the initial value evaluation. The maximum $\beta$ values are set to 50 for $\beta_d$ and 80 for $\beta_t$. $\beta$ for the overhang angle constraint is fixed at 50, while the overhang angle threshold $\bar{\alpha}$ is set to $45^\circ$ \cite{Guo2025, Zhao2020}. All examples considered in this paper are non-dimensionalized.
\subsection{L-shaped beam considering material isotropy}
\begin{figure*}[hbt!]
	\centering
	\includegraphics[scale=0.9]{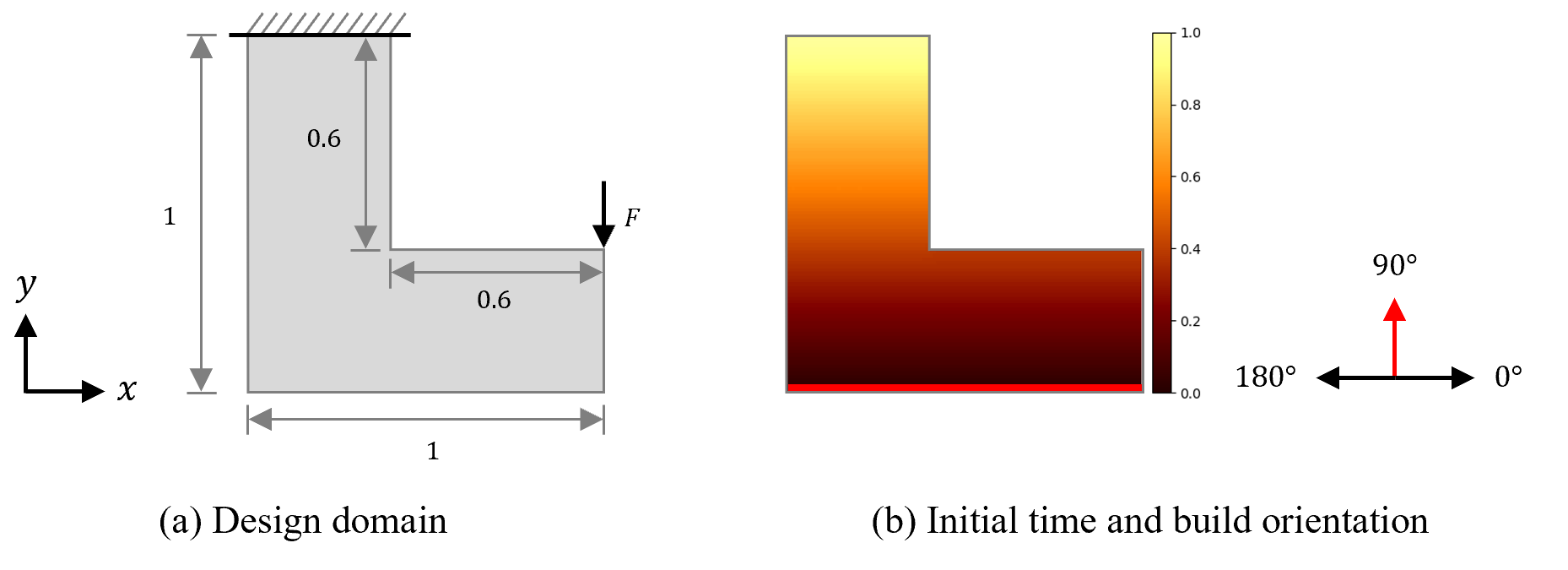}
	\caption{(a) Design configuration of the L-shaped beam and (b) initial time field and build orientations. The red line in the time field represents the base plate, where $\tau_e$ is fixed at 0, the red arrow represents the initial build orientations, and the black arrows denote the range of allowable build orientations.}
	\label{fig.7}
\end{figure*}
The design domain and boundary conditions of the problem are shown in Fig. \ref{fig.7}, where the optimal design is inherently expected to violate the overhang angle constraint \cite{Wang2020a}. The domain dimensions are presented in Fig. \ref{fig.7}(a), and the domain is uniformly discretized into square elements of length 0.01, resulting in a total of $6,400$ mesh elements. The initial density is set to 0.5 and is uniformly distributed throughout the domain, satisfying the material volume constraint. The initial time field $\bm \tau$ and build orientation $\bm\theta$ are illustrated in Fig. \ref{fig.7}(b), showing that their values are linearly dependent on the $y$ coordinate and range between 0 and 1. The initial build orientation is set perpendicular to the base plate, with a range of $\pm \pi / 2$. It is worth noting that the initial build orientation, in effect, determines the initial time distribution because the build orientation dictates the direction in which a structure is manufactured from the base plate. The performance is evaluated by adjusting the number of stages under the same conditions and comparing the results.
\begin{figure*}[hbt!]
	\centering
	\includegraphics[scale=0.9]{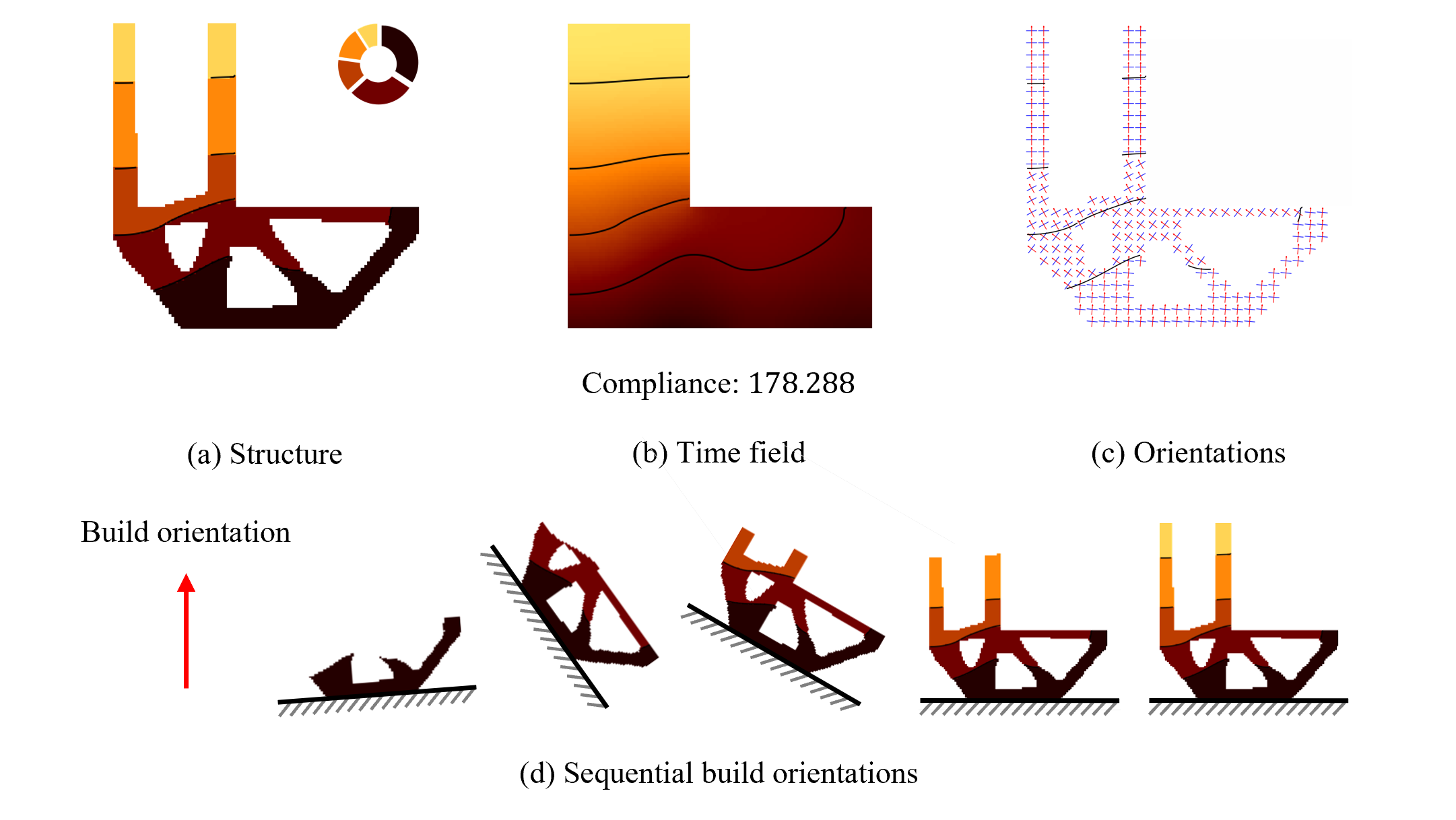}
	\caption{Optimized design configurations for the isotropic L-shaped beam with 5 stages: (a) optimized structure with stage-wise volume ratio, (b) time field, (c) orientations, and (d) sequential build orientations across 5 stages, considering stage-specific overhang angle constraint.}
	\label{fig.8}
\end{figure*}
\begin{table}[hbt!]
    \centering
    \begin{small}
    \caption{Stage-wise build orientation and volume ratio of an isotropic L-shaped beam with 5 stages.}
    \label{table.2}
    \begin{tabular}{@{}l l l l l l l@{}}
        \toprule
        Stage & Results & 1 & 2 & 3 & 4 & 5 \\  
        \midrule
        \multirow{2}{*}{5 stages} 
        & Build orientation & $84.92^\circ$ & $143.05^\circ$ & $119.09^\circ$ & $90.15^\circ$ & $90.01^\circ$ \\ 
        & Volume ratio   & $34.24\%$ & $28.96\%$ & $14.11\%$ & $13.55\%$ & $9.14\%$ \\ 
        \bottomrule
    \end{tabular}
    \end{small}
\end{table}

First, the optimized results for 5 stages (i.e., $N=5$) are summarized in Fig. \ref{fig.8}. Figure \ref{fig.8}(a) shows the optimal material layout, displaying the spatial material distribution $\bm{\rho}$ and a sequence of manufacturing stages, which are truncated from time field distribution $\bm{\tau}$ shown in Fig. \ref{fig.8}(b). It is also observed that the amount of material added in each sequence is nonuniform, as marked in the inset of Fig \ref{fig.8}(a), demonstrating the flexibility of the method. The build orientations $\bm\theta$ are plotted on the structure in Fig. \ref{fig.8}(c). To facilitate understanding, a simplified illustration is provided in Fig. \ref{fig.8}(d), depicting changes in build orientation and material deposition throughout the manufacturing sequence.
The designed structure is consistent with well-established numerical results where AM-oriented constraints are not considered \cite{Wang2020a}, while also satisfying overhang angle constraints at every manufacturing stage by leveraging flexible changes in build orientations over time. Such a finding manifests the significant improvement in manufacturability with minimal compromise in the performance, enabled by the proposed method.
Moreover, the time field successfully separates the stages while maintaining time continuity, ensuring that no isolated material or deposition occurs within already fabricated structures. This confirms that the framework satisfies the time continuity constraint. 
Detailed numerical results regarding stage-wise build orientation and volume ratio are summarized in Table \ref{table.2}. 
\begin{figure*}[hbt!]
	\centering
	\includegraphics[scale=1]{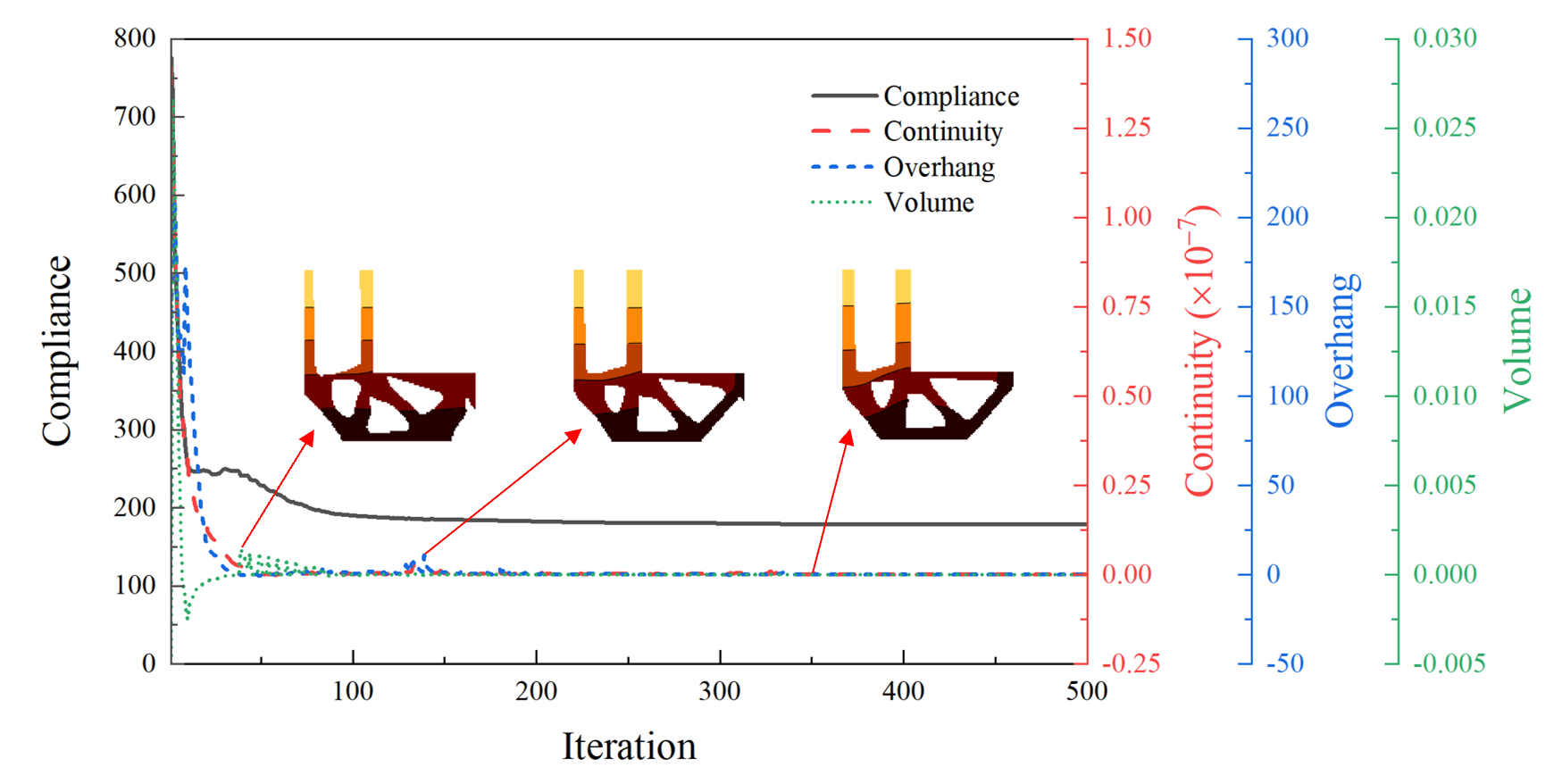}
	\caption{Convergence graph for the isotropic L-shaped beam with 5 stages: the black line represents compliance, the red dashed line represents the continuity constraint, the blue dashed line represents the overhang angle constraint, and the green dashed line represents the volume constraint.}
	\label{fig.9}
\end{figure*}

The convergence graph of the objective and constraint function values of the design optimization problem is shown in Fig. \ref{fig.9}. During the optimization process, oscillations in all functions were observed until the projection parameters $\beta_d$ and $\beta_t$ were increased. Beyond that point, the compliance value gradually converged to the optimal value while all constraints remained active. The fluctuations observed after the update of $\beta_d$ and $\beta_t$ are attributed to changes in build orientation. However, these fluctuations were minimal and had no noticeable impact on convergence. Only minor topological changes in the design were observed after iteration 350, indicating that the design was nearly stabilized. This suggests that the proposed framework can maintain stability while handling such variations.
\begin{figure*}[hbt!]
	\centering
	\includegraphics[scale=0.9]{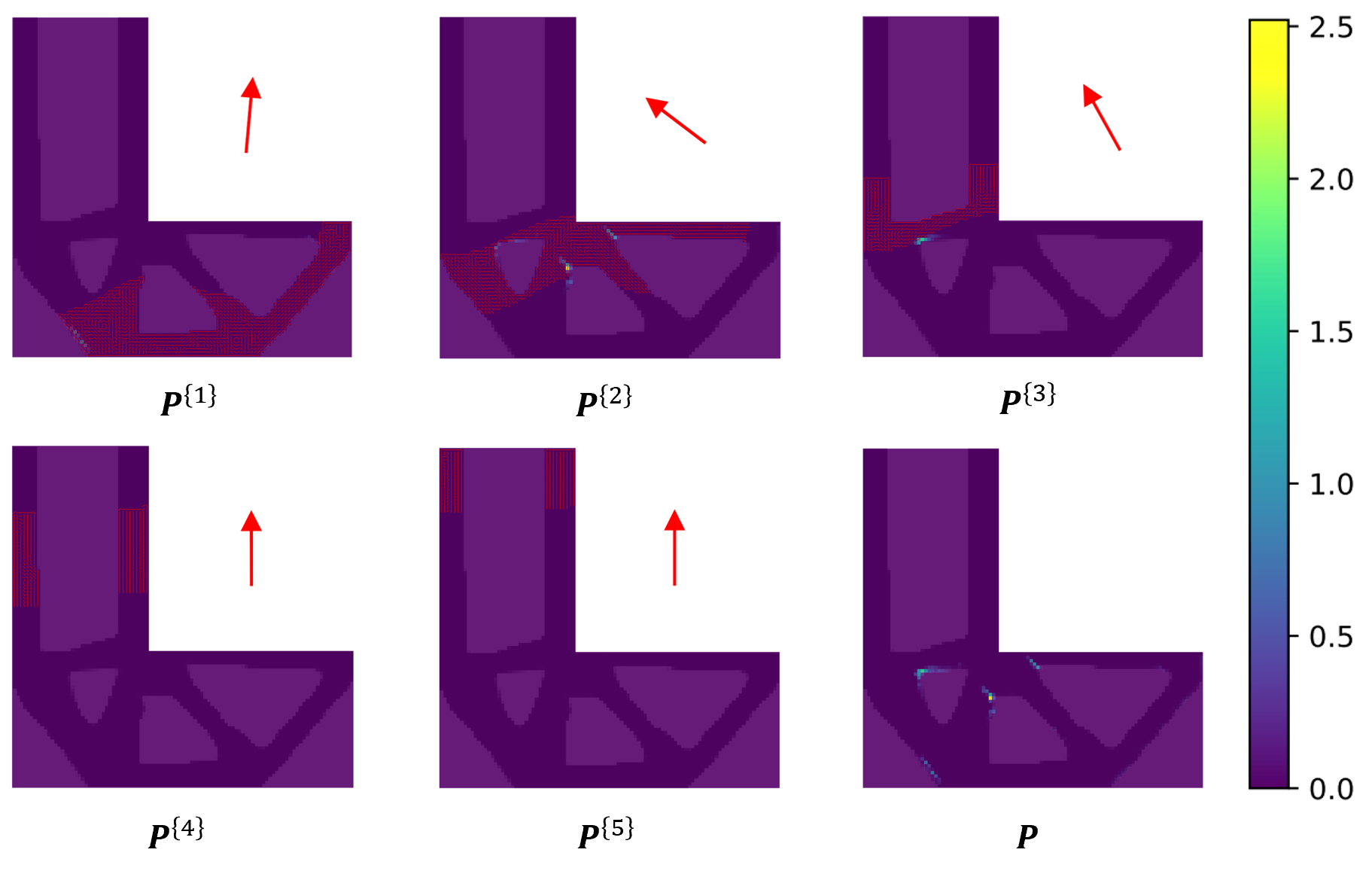}
	\caption{Overhang angle constraint values for each stage and cumulative values for the entire domain of the isotropic L-shaped beam with 5 stages. The red arrows indicate the build orientation for each stage, while the red lines within the elements represent the density boundaries of that stage.}
	\label{fig.10}
\end{figure*}

To further investigate the optimality of the solution, the overhang angle constraint values for each manufacturing stage $\bm{P}^{\{j\}}$ and $\bm{P}$ are examined as shown in Fig. \ref{fig.10}. As expected, since the overhang angle constraint is strictly enforced, the element-wise values are predominantly zero. First, the values are zero when there is no material (i.e., $\rho=0$) or when the material is not deposited during the corresponding stage, as predicted by Eq. (\ref{eq:overhang_constraint2}). Additionally, regions connected to the structure printed in the previous stage exhibit very low constraint values; previously fabricated structures act as self-supporting elements for subsequent stages, leaving them unaffected by overhang angle constraint. Nevertheless, it is worth noting that some regions exhibit slightly higher values of $P_e^{\{j\}}$ due in part to the aggregation method applied to $P_e^{\{j\}}$ \cite{Wang2020a}. For instance, $P_e^{\{j\}}$ reaches approximately 2.5 at the corner of the second stage of material deposition, a result of the fact that the corner surface is perpendicular to the build orientation. However, such constraint violation is negligible, as the affected elements at the corner constitute only a very small portion of the total structure, and their cumulative effect remains minimal. Overall, all areas satisfy the overhang angle threshold, confirming that the proposed framework effectively considers overhang angle constraint during optimization. 
\subsubsection[Effect of total number of stages N]{Effect of total number of stages $N$}
\begin{figure*}[hbt!]
	\centering
	\includegraphics[scale=0.9]{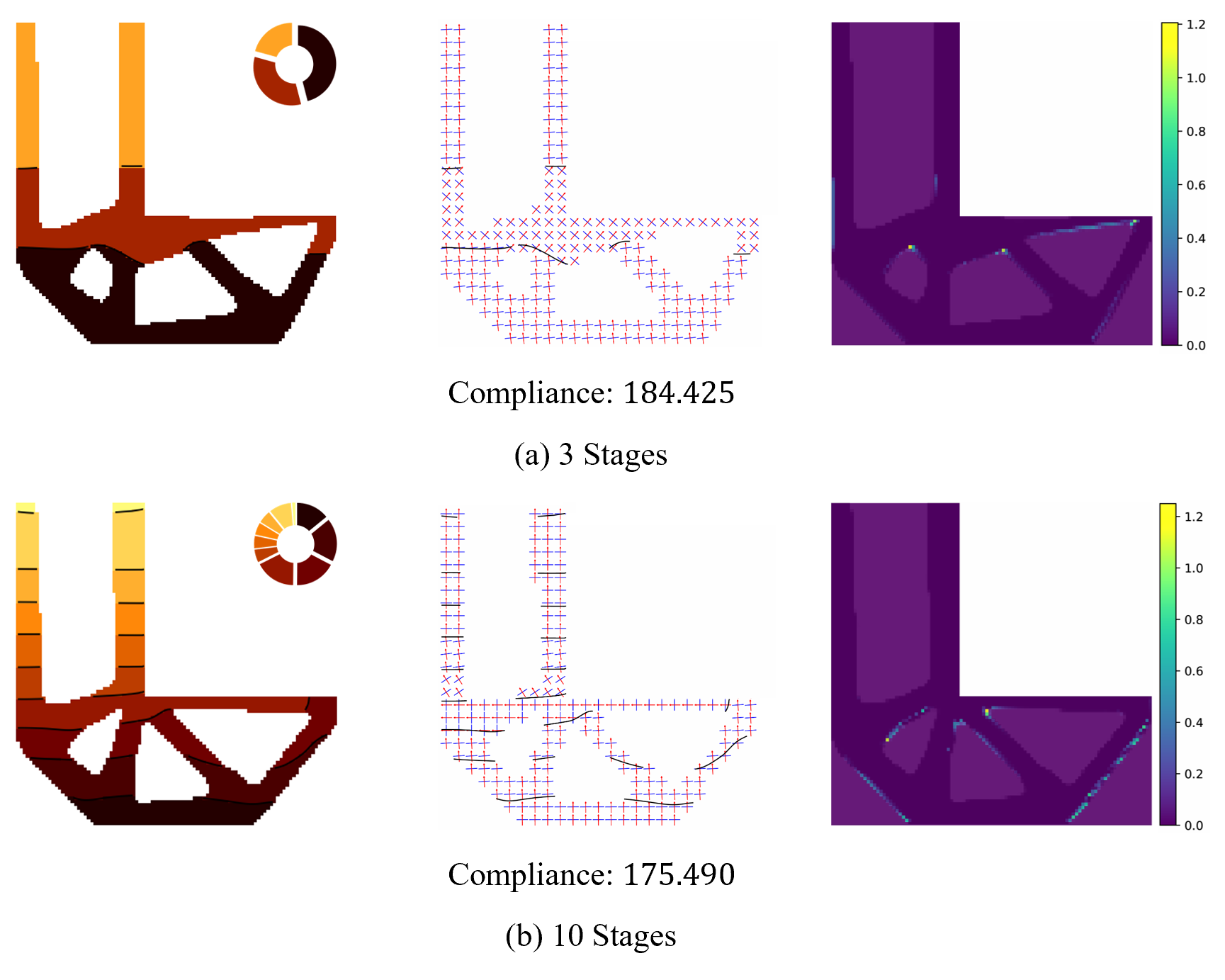}
	\caption{Optimal configurations for the isotropic L-shaped beam: (a) results with 3 stages, showing the optimized structure, stage-wise volume ratio, orientations, and cumulative overhang angle constraint values, and (b) results with 10 stages.}
	\label{fig.11}
\end{figure*}
To verify the applicability of the method across different numbers of stages and assess the optimality of solutions depending on the parameter, additional optimization problems are solved for cases where $N=3$ and $N=10$. The corresponding results are shown in Fig. \ref{fig.11} and Table \ref{table.3}, which display the deposition sequences of the optimized structure, their amounts of material, build orientation, and overhang values.

The results indicate that regardless of whether the number of stages is reduced to 3 or increased to 10, the density and time fields are optimized alongside changes in build orientations, ensuring compliance with the overhang angle threshold and resulting in a manufacturable structure.
When the specified number of stages is small ($N=3$), the optimized layouts deviate significantly from those in other cases where $N\geq5$, as shown in Fig. \ref{fig.11}(a). This deviation is due to the increased effect of the overhang angle constraint, which reduces design flexibility. Additionally, the optimized design exhibits higher compliance. In contrast, for higher values of $N$ as shown in Fig. \ref{fig.11}(b), the optimal design aligns with the reference design (i.e., $N=5$) and exhibits a slightly lower compliance value. The lower compliance value is attributed to the increased number of stages, which allows for diverse build orientations and reduces the influence of the overhang angle constraint, thus making the design process resemble an unconstrained case. However, it is worth noting that the build orientation converges to $90^\circ$ in stages beyond the 6th, as shown in Table \ref{table.3}, and the amount of material added in these steps is significantly lower, indicating that these additional stages are redundant. Since determining the number of stages requires manual adjustments, it becomes evident that a design method capable of adapting $N$ is required; nonetheless, this aspect is beyond the scope of the present study.
\begin{table}[hbt!]
    \centering
    \begin{small}
    \caption{Stage-wise build orientation and volume ratio for an isotropic L-shaped beam with 3 and 10 stages.}
    \label{table.3}
    \begin{tabular}{@{}l l l l l l l l l l l l @{}}
        \toprule
        Stage & Results & 1 & 2 & 3 & 4 & 5 & 6 & 7 & 8 & 9 & 10 \\ 
        \midrule
        \multirow{2}{*}{3 stages} 
        & Build orientation   &$96.39^\circ$ &$48.68^\circ$ &$93.07^\circ$ &-- &-- &-- &-- &-- &-- &-- \\ 
        & Volume ratio     &$46.04\%$ &$33.28\%$ &$20.68\%$ &-- &-- &-- &-- &-- &-- &-- \\ 
        \cmidrule(lr){1-12}
        \multirow{2}{*}{10 stages} 
        & Build orientation   &$90.56^\circ$ &$93.10^\circ$ &$93.96^\circ$ &$180.0^\circ$ &$123.66^\circ$ &$97.39^\circ$ &$90.0^\circ$ &$90.0^\circ$ &$90.0^\circ$ &$90.0^\circ$ \\ 
        & Volume ratio     &$14.12\%$ &$18.55\%$ &$17.58\%$ &$17.46\%$ &$5.50\%$ &$5.31\%$ &$5.22\%$ &$5.43\%$ &$9.46\%$ &$1.37\%$ \\ 
        \bottomrule
    \end{tabular}
    \end{small}
\end{table}
\begin{figure*}[hbt!]
	\centering
	\includegraphics[scale=0.9]{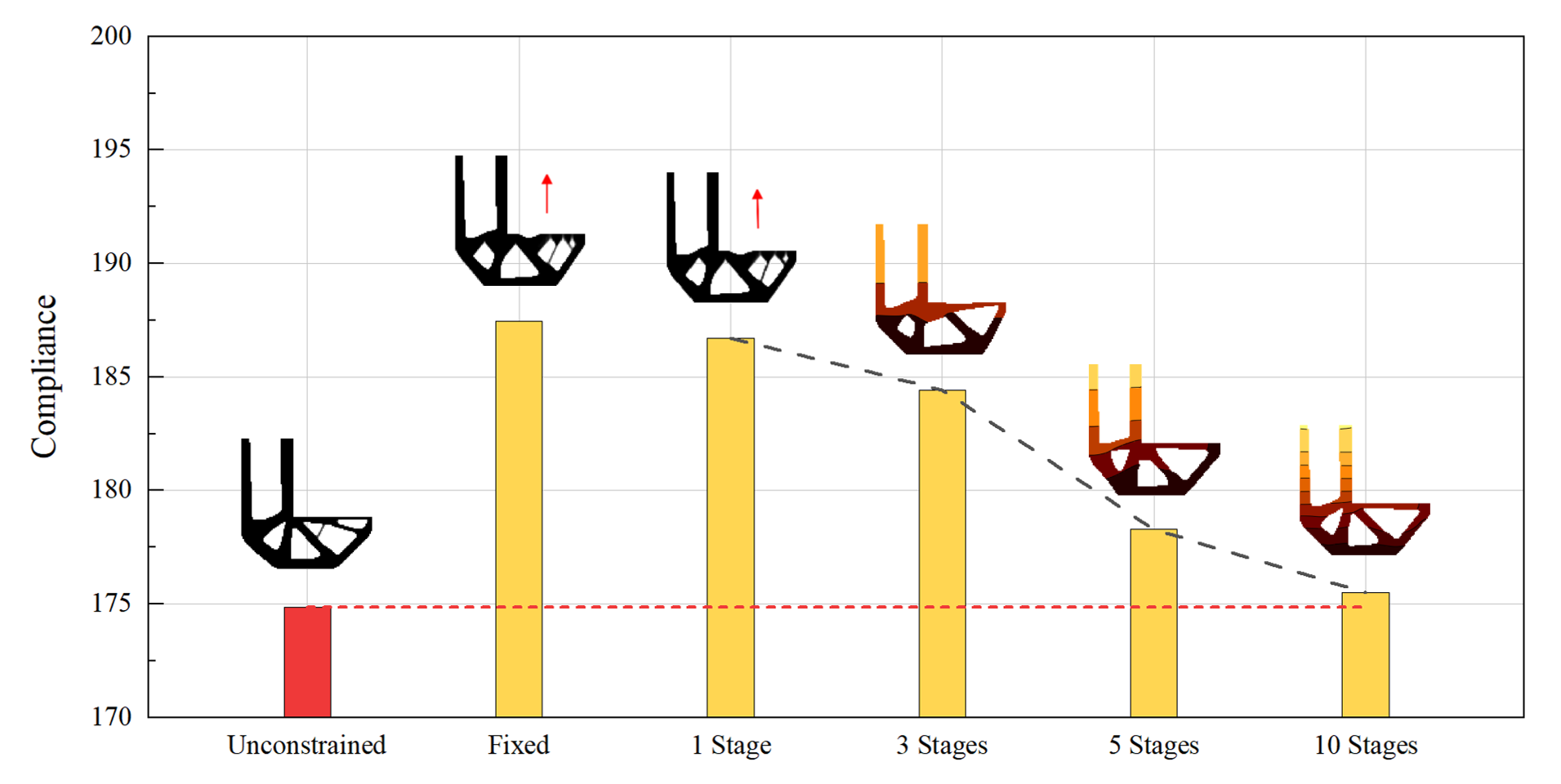}
	\caption{Comparison of optimized isotropic L-shaped beam models: the unconstrained model, the fixed build orientation model, and models with $N=1,3,5,10$. Red arrows indicate optimal build orientations, the red dashed line represents the compliance value of the unconstrained model, and the black dashed line shows the trend across the models.}
	\label{fig.12}
\end{figure*}

To further analyze how the optimized results change as $N$ increases, the design outcomes of the proposed method with $N=1,3,5,10$ are examined and compared with that of the unconstrained model, which only considers the material volume constraint, and the topology optimization model with a fixed build orientation under the overhang angle constraint \cite{Wang2020a}. Notably, OpenMDAO greatly facilitates flexibility in modifying the optimization configuration due to its reconfigurability. The optimized layouts, along with their corresponding compliance values under these conditions, are summarized in Fig. \ref{fig.12}.

Firstly, in the $N=1$ case, which represents a single build orientation, the adjustments of build orientation allow improvements in structural optimality compared to the fixed-build orientation ($\theta=90^\circ$), with the orientation resulting in a marginal change ($\theta=91.33^\circ$), as shown in Fig. \ref{fig.12}.

Furthermore, increasing the number of stages ($N=3,5,10$) provides greater flexibility in achieving manufacturable structures compared to the $N=1$ case by addressing the overhang angle constraint at each stage. As $N$ increases, the final structures increasingly resemble that of the unconstrained case. This trend is accompanied by compliance values that progressively converge toward that of the unconstrained model. Notably, the build orientation in the $N=10$ case becomes nearly fixed after the 6th stage, yet its objective function remains significantly smaller than that of the $N=5$ case. This improvement is due to the increased number of stages, which reduces the amount of deposited material in each of the earlier stages and allows for finer adjustments in build orientation, enabling more precise control. Consequently, these findings demonstrate that an increase in the number of stages enhances the resolution of the overhang angle constraint, leading to improved performance and manufacturable designs in multi-axis AM.
\subsubsection{Stage-specific and global material volume constraint}
In this section, the results obtained with a fixed amount of material deposition at each stage are examined, The stage-specific volume constraints are given by:
\begin{equation}
\begin{aligned}
    \sum_e \left( \rho_e^{\{j\}} - \rho_e^{\{j-1\}} \right) v_e - \frac{V_0}{N} \leq 0, \quad j = 1, 2, \dots, N
    \label{eq:stage_specific_constraint}
\end{aligned}
\end{equation}

\begin{figure*}[hbt!]
	\centering
	\includegraphics[scale=0.9]{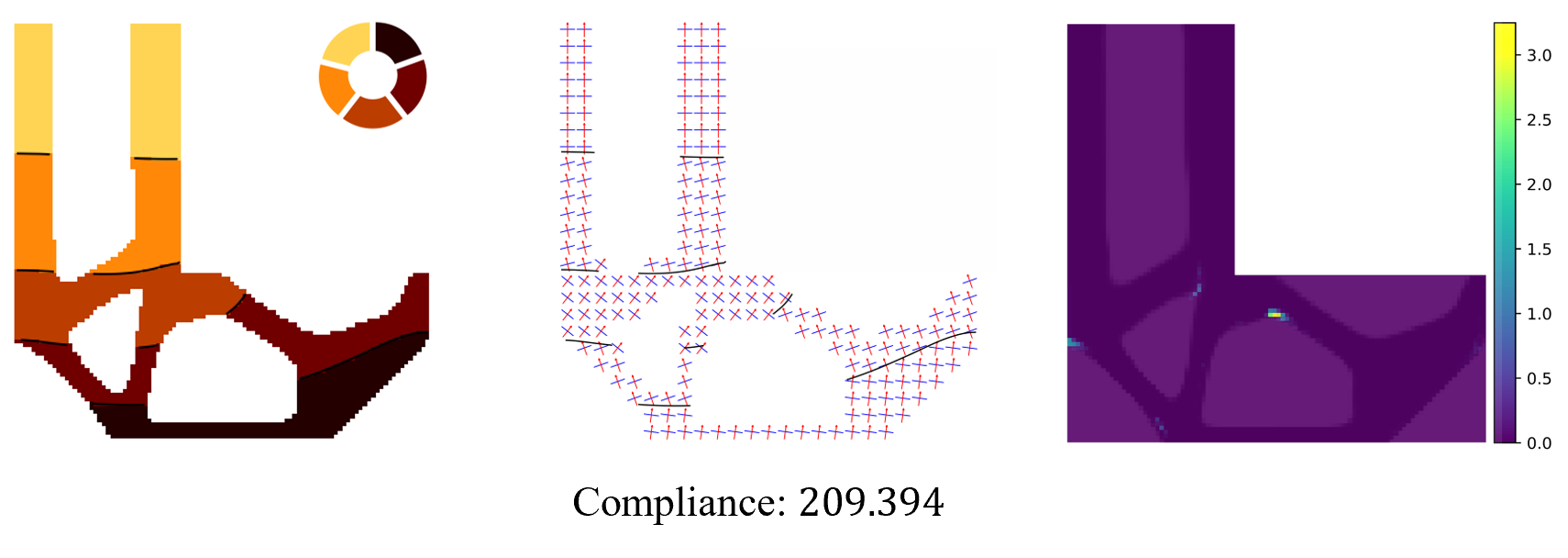}
	\caption{Optimal configurations for an isotropic L-shaped beam with 5 stages, considering stage-specific volume constraints: optimized structure, stage-wise volume ratio, orientations, and cumulative overhang angle constraint values.}
	\label{fig.13}
\end{figure*}
\begin{table}[hbt!]
    \centering
        \begin{small}
        \caption{Stage-wise build orientation of an isotropic L-shaped beam with 5 stages, considering stage-specific constraints.}
        \label{table.4}
        \begin{tabular}{@{}l l l l l l l @{}}
            \toprule
            \multirow{2}{*}{N=5}
            & Result & 1 & 2 & 3 & 4 & 5 \\ 
            \cmidrule(lr){2-7}
            & Build orientation   &$82.22^\circ$ &$109.46^\circ$ &$51.83^\circ$ &$104.50^\circ$ &$89.95^\circ$ \\ 
            \bottomrule
        \end{tabular}
        \end{small}
\end{table}

The equation (\ref{eq:stage_specific_constraint}) replaces the total material volume constraint given in Eq. (\ref{eq:volume_constraint}). The corresponding results for $N=5$ are presented in Fig. \ref{fig.13} and Table \ref{table.4}. These results confirm that the overhang angle constraint remains satisfied even under stage-specific volume constraints, producing manufacturable structures. However, the optimized material layout exhibits a significant discrepancy from the reference case without manufacturing constraints or the case with a global material volume constraint, as shown in Fig. \ref{fig.8}, leading to a higher compliance value.

The reduced optimality is attributed to differences in stage-wise build orientations. In the flexible volume constraint scenario, more than half (63.2\%) of the material is deposited during the first two stages in the bending-dominant part (i.e., the lower half) of the L-beam. A smaller amount of material than that of the remaining three stages (36.8\%) is used to create straight, slender bars in the tension-dominant region. However, when the stage-wise deposition is fixed, around 40\% of the total material volume, is deposited in the tension-dominant region during the last two stages, resulting in an unnecessarily bulky structure. Therefore, it can be concluded that an adaptable approach to stage-wise material deposition is essential in the present topology optimization method for AM.
\subsubsection[Effect of the continuity parameter gamma]{Effect of the continuity parameter $\gamma$}
During additive manufacturing, collisions can occur due to contact between the nozzle and already fabricated structures \cite{Guo2025,Lu2023,Ye2023}. Despite the importance of preventing such manufacturing issues, typical remedies require the post-process of the optimized design, potentially compromising its optimality. Nevertheless, directly addressing collision constraints in additive manufacturing, particularly in multi-axis systems, remains a significant challenge. Since collisions are inherently non-differentiable \cite{Guo2025}, incorporating this constraint necessitates computationally intensive methods such as finite differences. Otherwise, the optimality of the design is significantly degraded due to the limited sensitivity of the design variable. In this regard, this study proposes an alternative approach to indirectly manage collision constraints by modifying continuity constraints, thereby avoiding additional computational overhead.
\begin{figure*}[hbt!]
	\centering
	\includegraphics[scale=0.9]{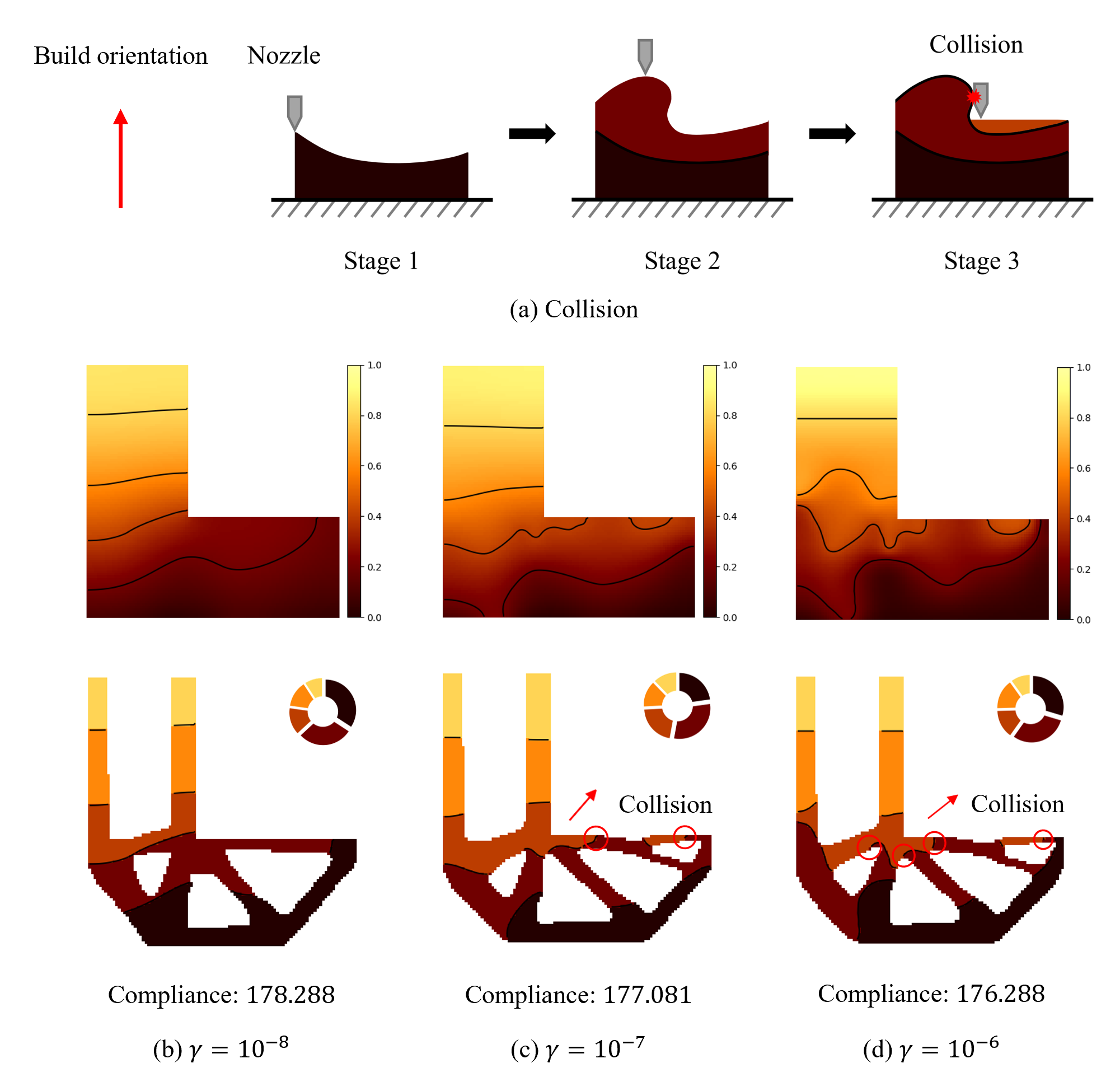}
	\caption{Collision in multi-axis AM: (a) an illustration showing collisions during the multi-axis AM process across stages and (b) the optimized time field, structure, and stage-wise volume ratio for different time continuity parameters in the isotropic L-shaped beam with 5 stages. The red arrow indicates the build orientation of the third stage, and the red circle highlights the collision point with the second stage.}
	\label{fig.14}
\end{figure*}
\begin{table}[hbt!]
    \centering
    \begin{small}
    \caption{Stage-wise build orientation and volume ratio of an isotropic L-shaped beam with 5 stages, considering varying continuity parameter values.}
    \label{table.5}
    \begin{tabular}{@{}l l l l l l l@{}}
        \toprule
        Continuity & Results & 1 & 2 & 3 & 4 & 5 \\ 
        \midrule
        \multirow{2}{*}{$\gamma = 10^{-8}$} 
        & Build orientation & $84.92^\circ$ & $143.05^\circ$ & $119.09^\circ$ & $90.15^\circ$ & $90.01^\circ$ \\ 
        & Volume ratio   & $34.24\%$     & $28.96\%$     & $14.11\%$     & $13.55\%$     & $9.14\%$ \\ 
        \midrule
        \multirow{2}{*}{$\gamma = 10^{-7}$} 
        & Build orientation & $70.29^\circ$ & $130.82^\circ$ & $49.06^\circ$ & $96.20^\circ$ & $90.0^\circ$ \\ 
        & Volume ratio   & $22.78\%$ & $30.06\%$ & $21.35\%$ & $13.58\%$ & $12.23\%$ \\ 
        \midrule
        \multirow{2}{*}{$\gamma = 10^{-6}$} 
        & Build orientation & $63.25^\circ$ & $141.52^\circ$ & $38.34^\circ$ & $81.47^\circ$ & $90.0^\circ$ \\ 
        & Volume ratio   & $29.60\%$ & $30.28\%$ & $14.70\%$ & $15.42\%$ & $9.99\%$ \\ 
        \bottomrule
    \end{tabular}
    \end{small}
\end{table}

In the proposed framework, collisions between the structure and the tool occur due to a lack of smoothness at stage boundaries, as shown in Fig. \ref{fig.14}(a). As discussed in Section 2, the smoothness of material between stages is determined by the time continuity constraint parameter $\gamma$. Higher $\gamma$ values increase the flexibility of stage boundaries but also raise the likelihood of collisions. Conversely, lower $\gamma$ values can limit design adaptability. Therefore, an appropriate $\gamma$ value must be selected to balance collision mitigation and structural performance. The results for different values of $\gamma=10^{-8}, 10^{-7}, 10^{-6}$ are presented in Fig. \ref{fig.14}(b), (c), and (d), respectively. When $\gamma \leq 10^{-8}$, the boundaries become overly constrained, causing the optimization process to converge to trivial solutions or fail to satisfy the constraint. Table \ref{table.5} also presents the stage-wise build orientations and material depositions, demonstrating the variations arising from different values of $\gamma$.

As shown in Fig. \ref{fig.14}(b-d), stage boundaries become less smooth as $\gamma$ increases, leading to a higher likelihood of collisions, as highlighted in red. For example, collisions occur between the tool in the third stage and the already fabricated second stage during fabrication along a third build orientation. However, higher $\gamma$ values also provide increased flexibility in stage boundaries and structural layouts, resulting in lower compliance values. These findings highlight the importance of selecting an appropriate $\gamma$ value during the optimization process to balance manufacturability and structural performance.
\subsection{L-shaped beam considering material anisotropy}
Changes in the build orientation are closely linked to the scan path, leading to anisotropy in the fabricated structure. This section further investigates the influence of stage-wise build orientation on material anisotropy, which is enabled by the proposed method. The design domain, initial time field, and build orientation settings are identical to those in the isotropic case, as illustrated in Fig. \ref{fig.7}. The anisotropic material model considered herein is presented in Table \ref{table.1}.
\begin{figure*}[hbt!]
	\centering
	\includegraphics[scale=0.9]{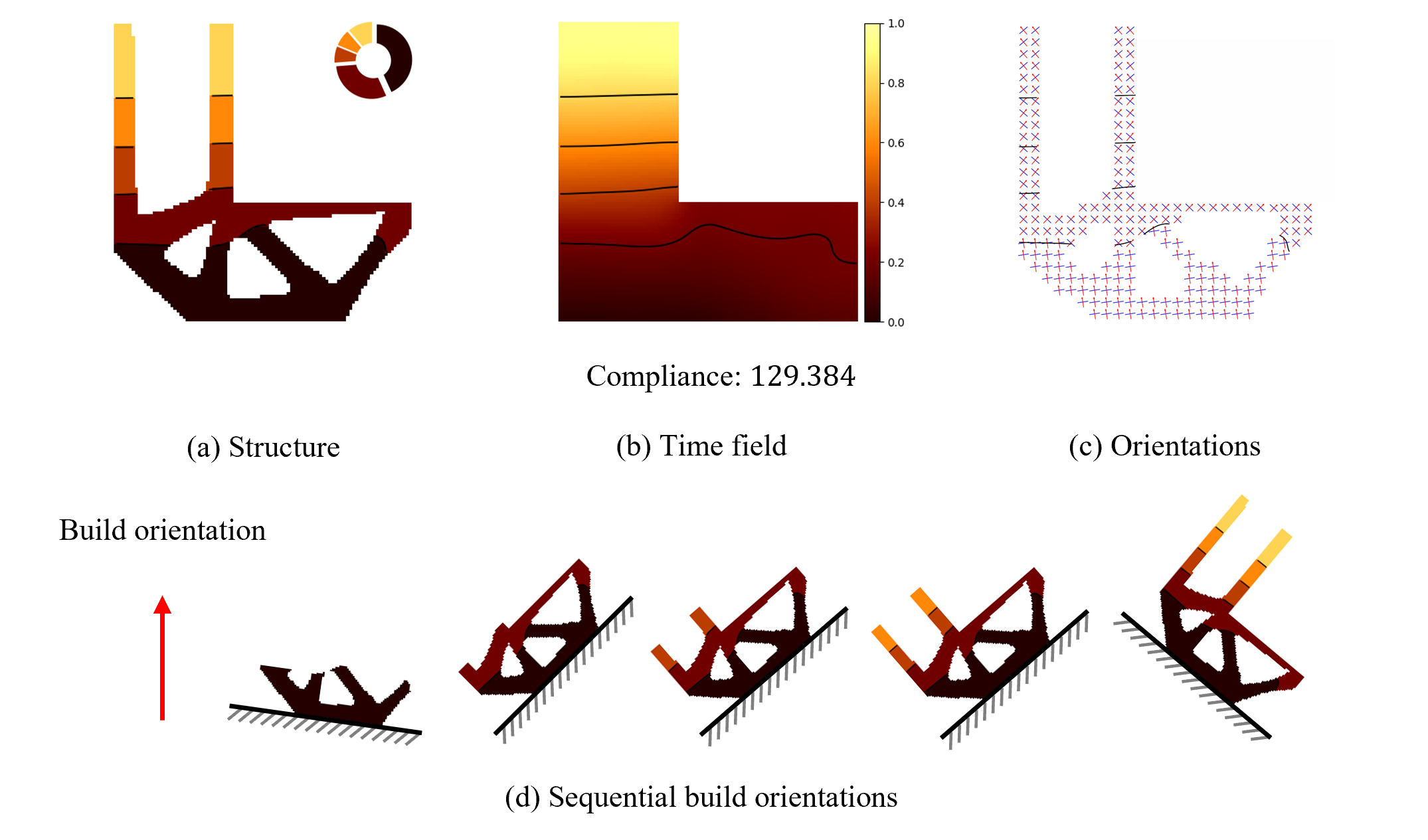}
	\caption{Optimized design configurations for the anisotropic L-shaped beam with 5 stages: (a) optimized structure with stage-wise volume ratio, (b) time field, (c) orientations, and (d) sequential build orientations across 5 stages, considering the stage-specific overhang angle constraint.}
	\label{fig.15}
\end{figure*}
\begin{table}[hbt!]
    \centering
    \begin{small}
    \caption{Stage-wise build orientation and volume ratio of an anisotropic L-shaped beam with 5 stages.}
    \label{table.6}
    \begin{tabular}{@{}l l l l l l l@{}}
        \toprule
        Stage & Results & 1 & 2 & 3 & 4 & 5 \\ 
        \midrule
        \multirow{2}{*}{5 stages} 
        & Build orientation   &$98.92^\circ$ &$45.99^\circ$ &$49.39^\circ$ &$49.56^\circ$ &$130.80^\circ$ \\ 
        & Volume ratio     &$43.20\%$ &$30.58\%$ &$7.42\%$ &$7.48\%$ &$11.32\%$ \\ 
        \bottomrule
    \end{tabular}
    \end{small}
\end{table}

The results for the anisotropic L-shaped beam with 5 stages are presented in Fig. \ref{fig.15} and Table \ref{table.6}. As shown in Fig. \ref{fig.15}(a-c), the stage-specific material deposition is tailored to construct the final structure, similar to the isotropic case. Both the time field and build orientations are optimized while satisfying the given continuity and overhang angle constraint. However, unlike in the isotropic case, most parts of the structure are fabricated at angles close to $45^\circ$ and $135^\circ$. This trend is attributed to the orthotropic properties of the considered material, which maximize tensile stiffness in these directions, thereby enhancing structural stiffness. Furthermore, compliance is lower in the anisotropic case compared to the isotropic case, since the maximum Young’s modulus of the isotropic material is 1, whereas the anisotropic material reaches approximately 1.5 along the principal axis (Fig. \ref{fig.4}(c)). Fig. \ref{fig.15}(d) further illustrates that the structure is optimized to align with these orientations, minimizing compliance.

An analysis of the stage-specific volume ratio reveals that approximately $73.8\%$ of the material is deposited within the first two stages for the anisotropic beam, compared to $63.2\%$ in the isotropic model. This difference arises from the adaptation of the time field to material property variations introduced by stage-specific build orientations, which are then integrated into the structure. Additionally, this observation suggests that the straight beams in tension-dominant regions become thinner, highlighting the ability of the proposed framework to optimize structural design in response to material anisotropy. These results highlight the effectiveness of the proposed framework in incorporating manufacturing constraints and material anisotropy in the optimization process for multi-axis AM.
\begin{figure*}[hbt!]
	\centering
	\includegraphics[scale=1]{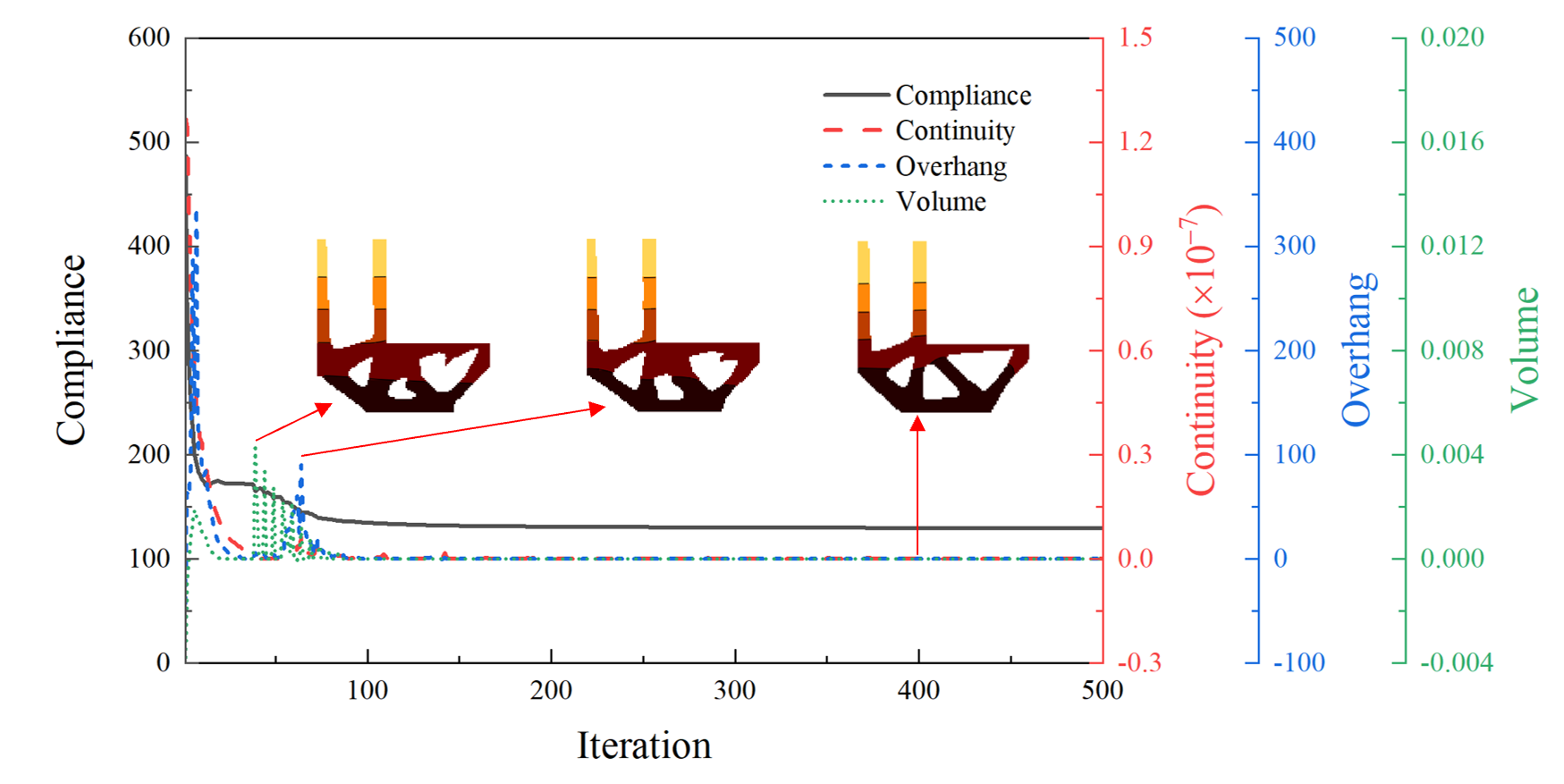}
	\caption{Convergence graph for the anisotropic L-shaped beam with 5 stages: the black line represents compliance, the red dashed line represents the continuity constraint, the blue dashed line represents the overhang angle constraint, and the green dashed line represents the volume constraint.}
	\label{fig.16}
\end{figure*}
\begin{figure*}[hbt!]
	\centering
	\includegraphics[scale=0.9]{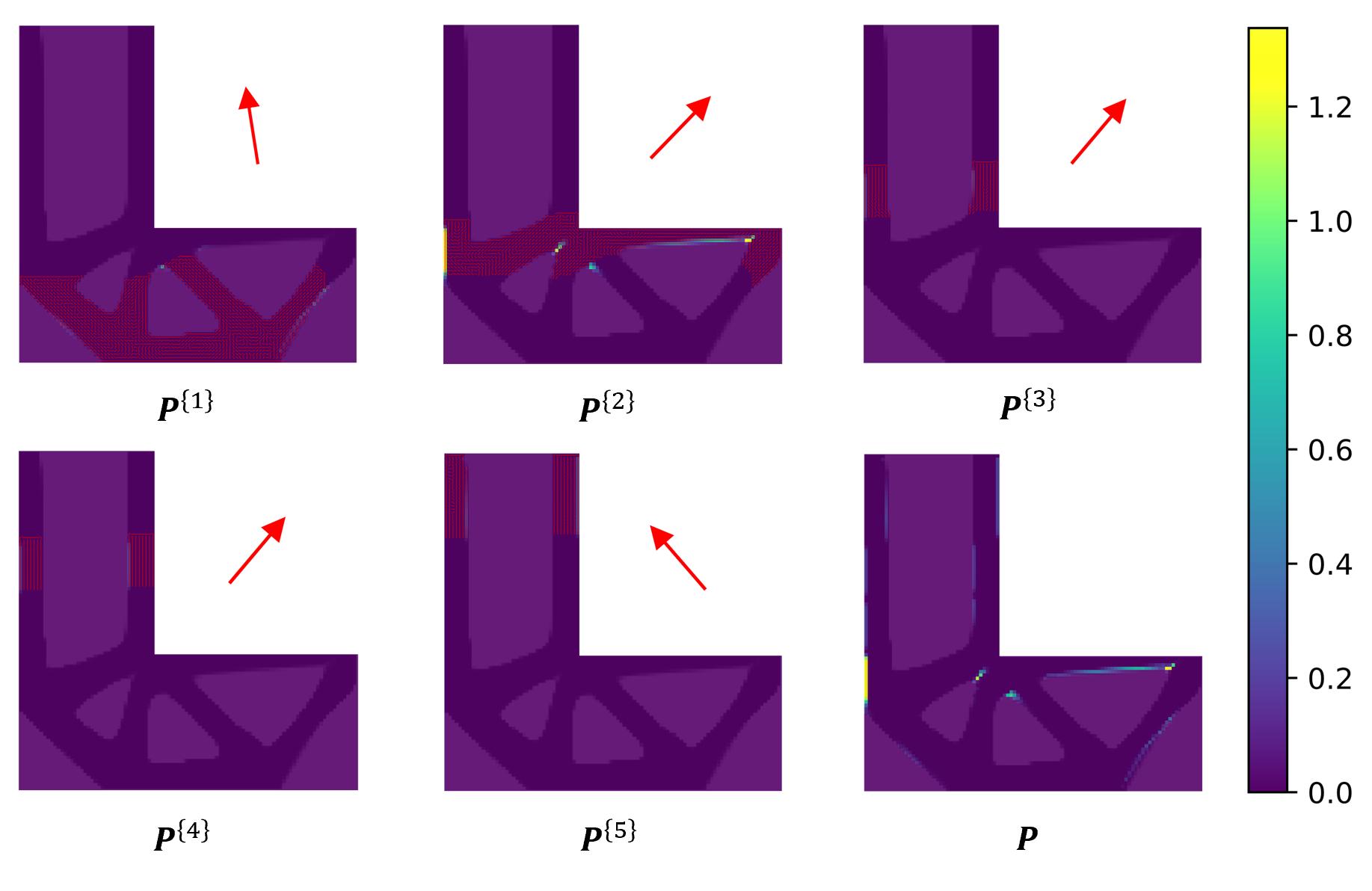}
	\caption{Overhang angle constraint values for each stage and the cumulative values for the entire domain of the anisotropic L-shaped beam with 5 stages. The red arrows indicate the build orientation for each stage, while the red lines within the elements represent the density boundaries of that stage.}
	\label{fig.17}
\end{figure*}

The convergence graph for the anisotropic L-shaped beam with 5 stages is shown in Fig. \ref{fig.16}. 
It is found that the overall convergence is similar to the isotropic case, except that relatively large constraint violations were observed in the early stages of optimization. The oscillations were predominantly observed between iterations 40 to 140, during which the projection $\beta$ value was periodically updated. The increased oscillations can be attributed to the wider variance in build orientations in the anisotropic case, rendering the overhang angle constraint harder to satisfy. Consequently, material is reintroduced in regions where it was previously removed, such as around iteration 60, as shown in Fig. \ref{fig.16}. Nevertheless, as the iteration progresses, the optimization process stabilizes, and the solution gradually converges to a feasible design.

Figure \ref{fig.17} illustrates the stage-wise overhang angle constraint $\bm{P}^{\{j\}}$ for the intermediate structures and the overall constraint $\bm{P}$, which combines contributions from all stages. Compared to the isotropic case, the overhang angle constraint values are relatively higher at the structural boundaries. Nevertheless, all these values remain within the overhang angle threshold of $45^\circ$. This phenomenon arises because build orientations in the anisotropic material model are influenced not only by the overhang angle constraint but also by material properties. Consequently, build orientations close to the overhang threshold of $45^\circ$ are obtained to minimize the objective function. While this issue could potentially be addressed through the stricter aggregation of constraints (e.g., P-norm), it is not considered in the present study.
\subsubsection[Effect of total number of stages N]{Effect of total number of stages $N$}
\begin{figure*}[hbt!]
	\centering
	\includegraphics[scale=0.9]{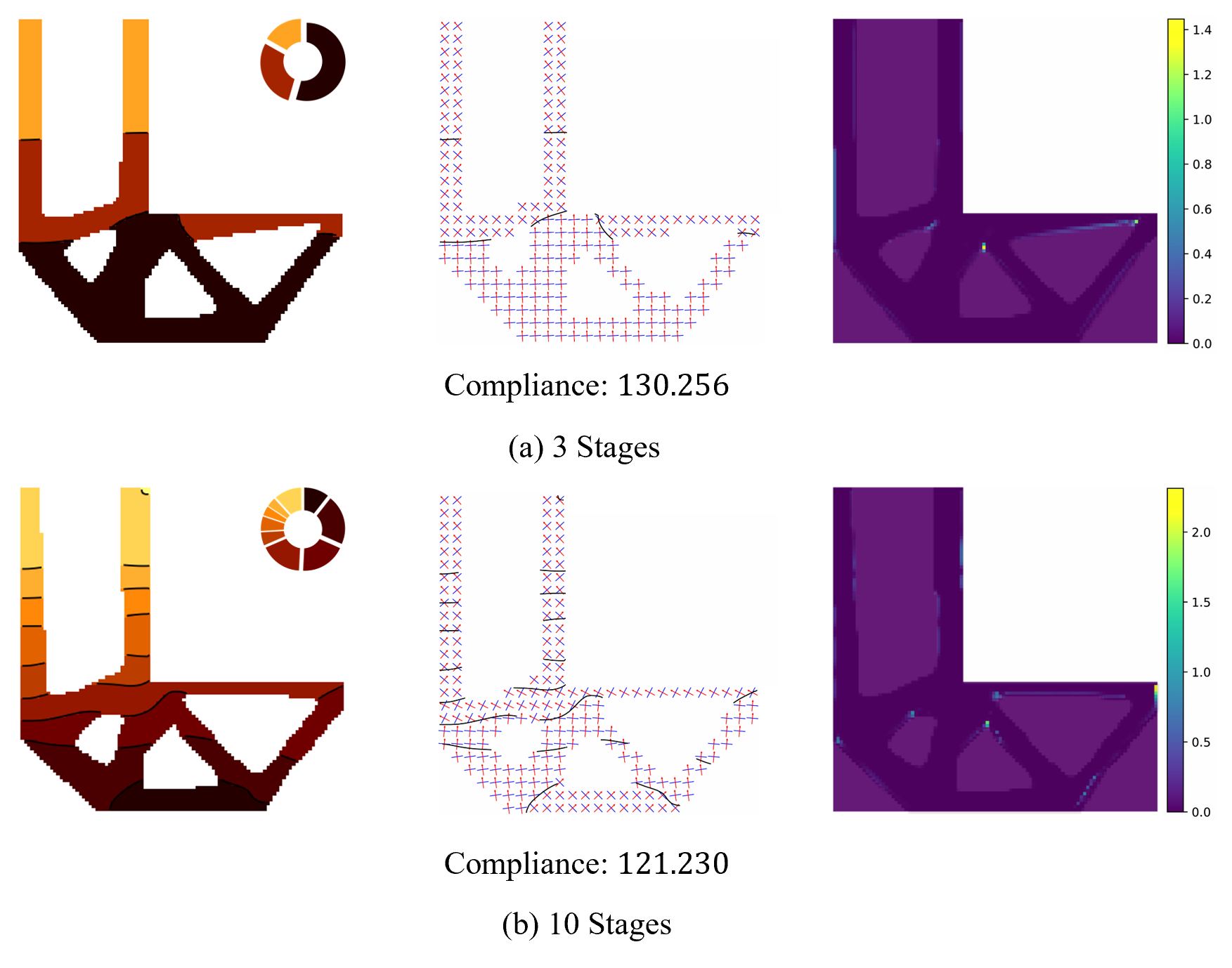}
	\caption{Optimal configurations for the anisotropic L-shaped beam: (a) results with 3 stages, showing the optimized structure, stage-wise volume ratio, orientations, and cumulative overhang angle constraint values, and (b) results with 10 stages.}
	\label{fig.18}
\end{figure*}
\begin{table}[hbt!]
    \centering
    \begin{small}
    \caption{Stage-wise build orientation and volume ratio for an anisotropic L-shaped beam with 3 and 10 stages.}
    \label{table.7}
    \begin{tabular}{@{}l l l l l l l l l l l l @{}}
        \toprule
        Stage & Results & 1 & 2 & 3 & 4 & 5 & 6 & 7 & 8 & 9 & 10 \\ 
        \midrule
        \multirow{2}{*}{3 stages} 
        & Build orientation   &$93.37^\circ$ &$48.42^\circ$ &$129.44^\circ$ &-- &-- &-- &-- &-- &-- &-- \\ 
        & Volume ratio     &$54.50\%$ &$28.69\%$ &$16.81\%$ &-- &-- &-- &-- &-- &-- &-- \\ 
        \cmidrule(lr){1-12}
        \multirow{2}{*}{10 stages} 
        & Build orientation   &$40.40^\circ$ &$98.31^\circ$ &$84.82^\circ$ &$152.46^\circ$ &$49.24^\circ$ &$49.17^\circ$  &$49.01^\circ$ &$131.77^\circ$ &$132.55^\circ$ &$130.63^\circ$ \\ 
        & Volume ratio     &$10.56\%$ &$21.08\%$ &$19.17\%$ &$17.71\%$ &$5.62\%$ &$6.12\%$ &$4.03\%$ &$3.93\%$ &$11.62\%$ &$0.16\%$ \\ 
        \bottomrule
    \end{tabular}
    \end{small}
\end{table}
To assess the effectiveness of the anisotropy model across different numbers of stages, optimization was performed for cases with $N=3$ and $N=10$. The results are shown in Fig. \ref{fig.18} and Table \ref{table.7}. Similar to the model with $N=5$, the anisotropy model demonstrates that the density, time field, and build orientation dynamically adapt to reflect the final structure, regardless of the number of stages.

Consistent with the observations for $N=5$, build orientations are predominantly aligned to $45^\circ$ and $135^\circ$. However, the convergence of the build orientation at the later stages, which was clearly observed in the isotropic case, did not occur even when the number of stages increased to $N=10$. Nevertheless, regardless of $N$, it is worth noting that the amount of material deposited in the shear-dominant region remains greater than that in the tension-dominant region, showcasing the consistency of the proposed method even when the material anisotropy is considered.
\begin{figure*}[hbt!]
	\centering
	\includegraphics[scale=0.9]{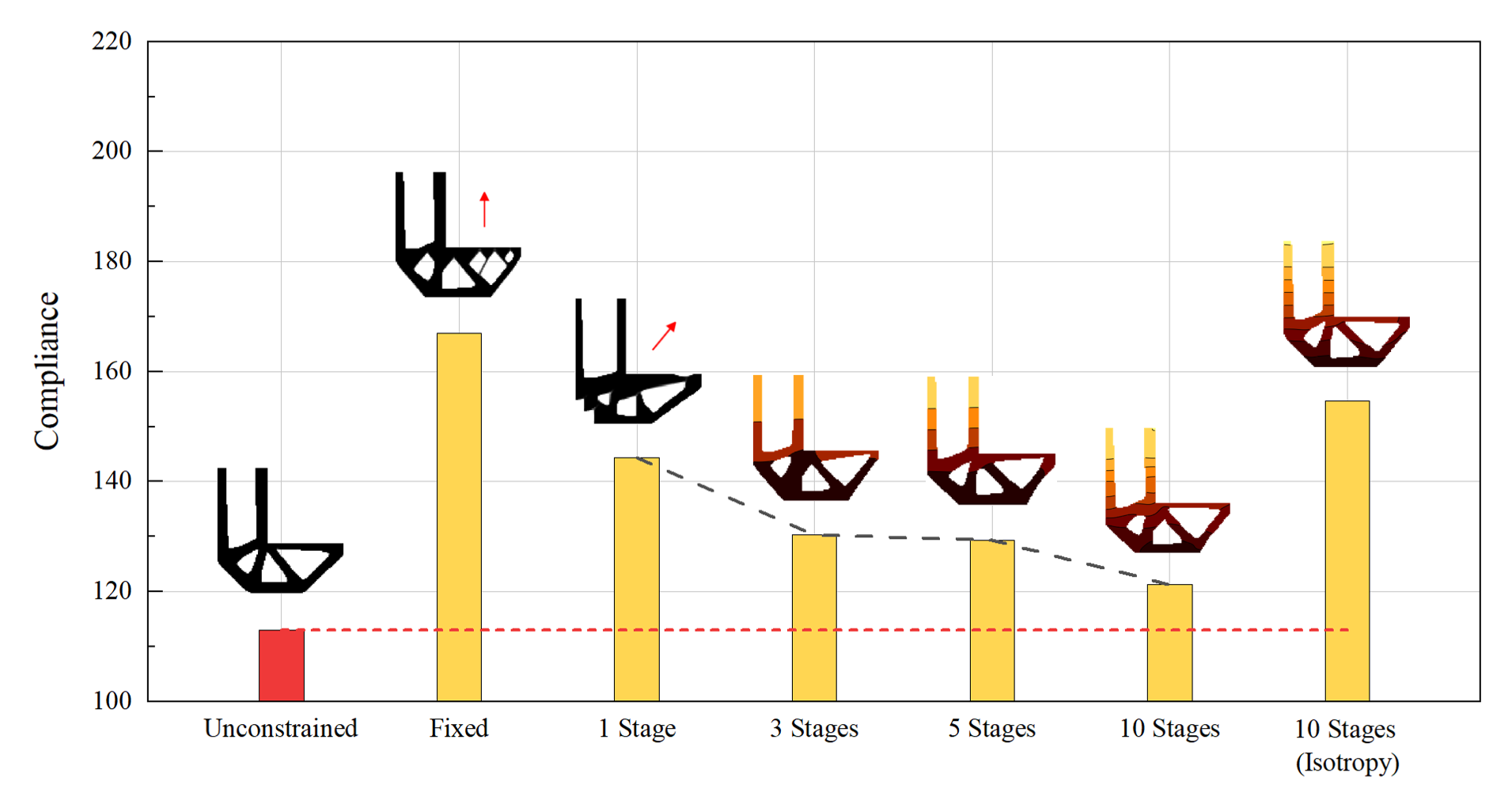}
	\caption{Comparison of optimized anisotropic L-shaped beam models: the unconstrained model, the fixed build orientation model, models with $N=1,3,5,10$, and the isotropic model with $N=10$. Red arrows indicate optimal build orientations, the red dashed line represents the compliance value of the unconstrained model, and the black dashed line shows the trend across the models.}
	\label{fig.19}
\end{figure*}
Figure \ref{fig.19} demonstrates the efficacy of the considered anisotropy. For both the fixed and $N=1$ cases, performance is less favorable than in the unconstrained case, which represents an ideal scenario where density and material orientation are independently optimized for each element, similar to the isotropic case. However, this deviation is more pronounced in anisotropic cases due to the dependence of material properties on build orientation. It is worth mentioning that the $N=1$ case achieves a significantly lower compliance value than the fixed orientation case, as the build orientation changes substantially to $\theta=50.67^\circ$, but this adjustment results in an invalid configuration requiring additional support structures. This result indicates that the anisotropic case prioritizes adjusting the build orientation to optimize material properties, followed by modifying the local material layout to satisfy the overhang angle constraint. Without defining a specific region as a base plate, such configurations could potentially lead to valid structures that do not require additional support, but this aspect was not addressed in the present study.

The multi-stage case ($N>1$) demonstrates significant performance improvements over the single-build orientation case (i.e., Fixed). As the number of stages increases and build orientations become more varied, the influence of the overhang angle constraint is reduced, causing the final structure and structural properties to converge more closely to those of the unconstrained case. Furthermore, when the results of the isotropic case with $N=10$ are reanalyzed by considering material orientation based on build orientation, a significant compliance difference is observed compared to the $N=10$ case with material anisotropy. This further demonstrates that the proposed framework effectively integrates anisotropic characteristics into the optimization process.
\subsection{Cantilever beam with cutout}
\begin{figure*}[hbt!]
	\centering
	\includegraphics[scale=0.9]{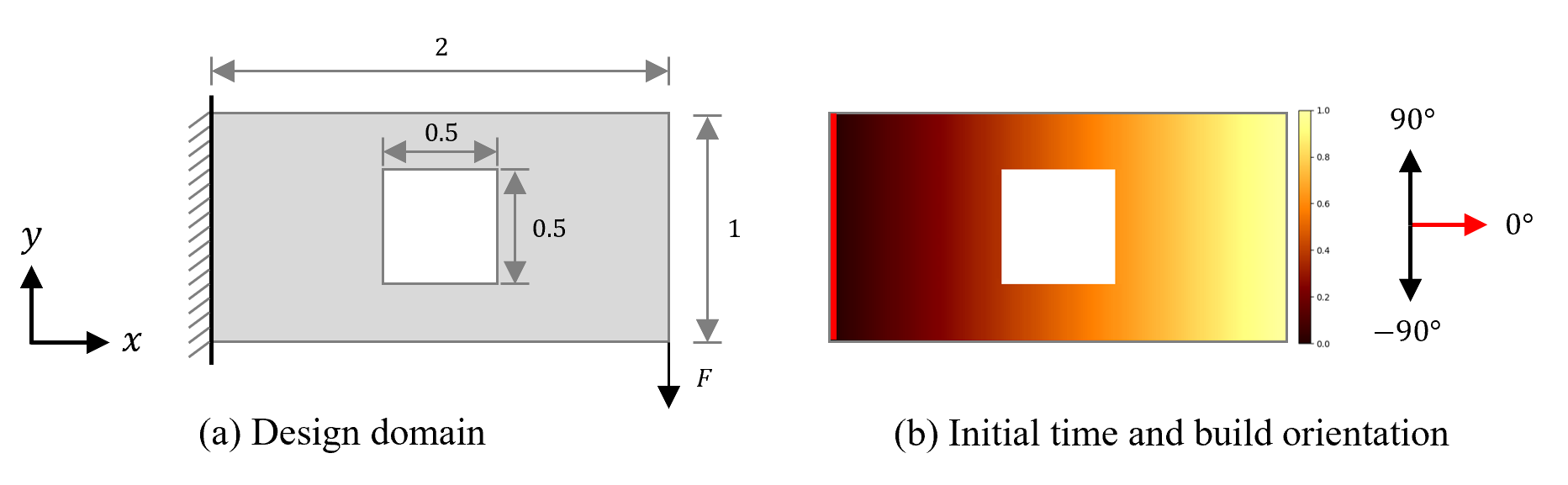}
	\caption{(a) Design configuration of the cantilever beam with cutout and (b) initial time field and build orientations. The red line in the time field indicates the base plate where $\tau_e$ is fixed at 0, the red arrow represents the initial build orientations, and the black arrows denote the range of allowable build orientations.}
    \label{fig.20}
\end{figure*}
This example considers a cantilever beam with a fixed cutout, as illustrated in Fig. \ref{fig.20} \cite{Wang2020a}. The design domain includes inherent overhangs, and the overhang angle constraint is violated from the beginning of the iteration. The design domain is discretized into 17,500 elements, and a force of $F=1$ is applied. The initial density is set to 0.5, while the initial time field and build orientations are shown in Fig. \ref{fig.20}(b).
\begin{figure*}[hbt!]
	\centering
	\includegraphics[scale=0.9]{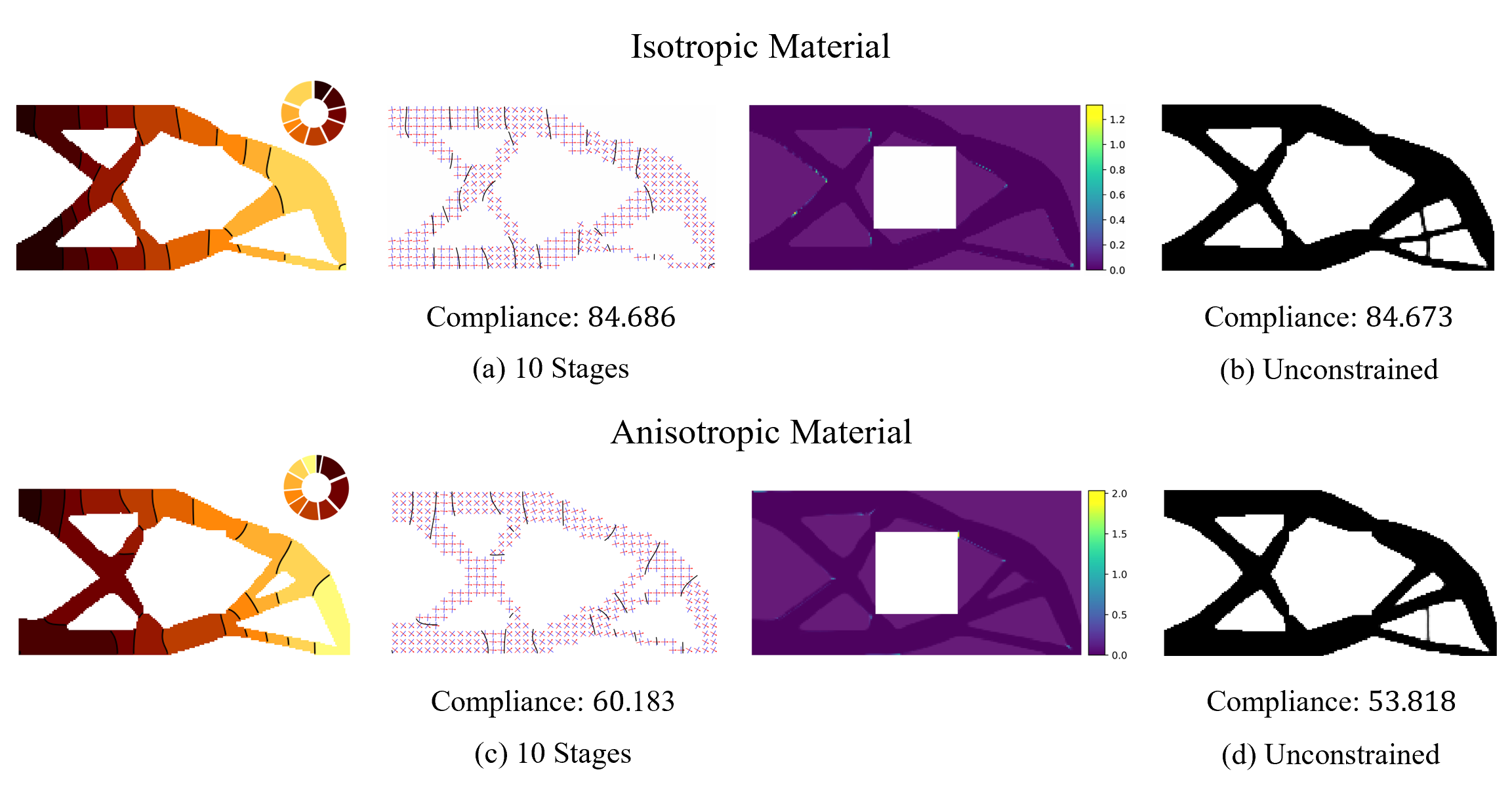}
	\caption{Optimal configurations for the cantilever beam with cutout: (a) results for the isotropic case with 10 stages, showing the optimized structure, stage-wise volume ratio, orientations, and cumulative overhang angle constraint values, (b) isotropic unconstrained model, (c) results for the anisotropic case with 10 stages, and (d) anisotropic unconstrained model.}
	\label{fig.21}
\end{figure*}
\begin{table}[hbt!]
    \centering
    \begin{small}
    \caption{Stage-wise build orientation and volume ratio of isotropic and anisotropic cantilever beams with cutout for a model with 10 stages.}
    \label{table.8}
    \begin{tabular}{@{}l l l l l l l l l l l l @{}}
        \toprule
        Type & Results & 1 & 2 & 3 & 4 & 5 & 6 & 7 & 8 & 9 & 10 \\ 
        \midrule
        \multirow{2}{*}{Isotropy} 
        & Build orientation   &$8.52^\circ$ &$-3.05^\circ$ &$3.15^\circ$ &$38.84^\circ$ &$-38.16^\circ$ &$0.89^\circ$ &$-31.49^\circ$ &$5.96^\circ$ &$-40.44^\circ$ &$-40.70^\circ$ \\ 
        & Volume ratio     &$9.83\%$ &$11.85\%$ &$8.82\%$ &$12.19\%$ &$11.90\%$ &$9.66\%$ &$5.21\%$ &$11.02\%$ &$19.36\%$ &$0.16\%$ \\ 
        \cmidrule(lr){1-12}
        \multirow{2}{*}{Anisotropy} 
        & Build orientation   &$-44.53^\circ$ &$-40.69^\circ$ &$-3.84^\circ$ &$42.07^\circ$ &$53.40^\circ$ &$24.30^\circ$ &$19.73^\circ$ &$14.11^\circ$ &$0.40^\circ$ &$-29.89^\circ$ \\ 
        & Volume ratio     &$2.84\%$ &$15.76\%$ &$19.33\%$ &$10.48\%$ &$10.96\%$ &$6.66\%$ &$7.47\%$ &$9.71\%$ &$9.14\%$ &$7.65\%$ \\
        \bottomrule
    \end{tabular}
    \end{small}
\end{table}

The optimization results with $N=10$ are summarized in Fig. \ref{fig.21} and Table \ref{table.8}, which depict material layouts, build orientations, and overhang values for both isotropic (Fig. \ref{fig.21}(a)) and anisotropic (Fig. \ref{fig.21}(c)) cases. The unconstrained cases for each material assumption are presented as reference cases in Fig. \ref{fig.21}(b) and Fig. \ref{fig.21}(d).

The results indicate that, despite the initial violations of the overhang angle constraint caused by the cutout, the proposed method redistributes material to mitigate these violations near the cutout region, demonstrating its applicability to various geometric configurations. As shown in Fig. \ref{fig.21}(a) and Fig. \ref{fig.21}(c), no significant topological differences are observed relative to the unconstrained cases, likely due to the high number of stages ($N=10$). However, as shown in Fig. \ref{fig.21}(b) and Fig. \ref{fig.21}(d), thin features that violate the overhang angle constraint in the unconstrained cases are eliminated, confirming the enforcement of the overhang angle constraint throughout the design domain.

For the isotropic case, compliance values are nearly identical to that of the unconstrained model, consistent with the observations from the L-shaped beam example shown in Fig. \ref{fig.12}. This again indicates that increasing the number of stages enables manufacturability constraints to be satisfied while preserving structural behavior close to that of an unconstrained design for isotropic materials. In contrast, compliance deviations from the unconstrained model are observed for the anisotropic case resulting from the combined effects of the overhang angle constraint and material properties which depend on build orientation, consistent with findings from the example in Fig. \ref{fig.19}.

\section{Conclusion} \label{sec:Conclusion}
This paper presents a novel space-time topology optimization framework for multi-axis AM. Using a pseudo-time field, the density distribution is divided into multiple stages, and the build orientation for each stage is simultaneously optimized under an overhang angle constraint. Additionally, anisotropic properties, corresponding to a material orientation perpendicular to the build orientation, are considered in the optimization process. Numerical examples demonstrate that both isotropic and anisotropic models produce feasible configurations and build orientations. Furthermore, increasing the number of stages leads to enhanced performance, making it comparable to that of the unconstrained model. Comparisons of different approaches to volume constraint application show that stage-specific volume adjustments, dynamically adapting to each stage, contribute to improved performance. In addition, adjusting the time continuity constraint parameter prevents potential collisions during the manufacturing process, further enhancing the feasibility of the proposed framework. These results confirm the capability of our approach to capture the diverse build orientations and anisotropic material properties of multi-axis AM during the optimization process. Furthermore, the findings emphasize how multi-stage build orientation strategies and the integration of manufacturing features into topology optimization collectively contribute to achieving superior performance.

Although the proposed framework effectively integrates build orientation optimization and anisotropic material properties, some limitations remain and require further investigation. One limitation is the overhang angle constraint, which is currently imposed using an averaged value rather than other constraint aggregation methods (e.g., p-norm). This choice was made to maintain consistency with previous studies \cite{Wang2020a}, but it does not strictly enforce the constraint and may require refinement for improved accuracy. Additionally, while the proposed collision prevention strategy significantly mitigates potential collisions, it lacks intuitive interpretability and could benefit from further improvements. Extending the framework to 3D geometries also presents challenges, particularly in capturing material anisotropy, as the numerous deposition paths in each layer influence structural performance.

To address these challenges, future research will explore methods to explicitly enforce overhang angle constraints and develop more intuitive collision prevention strategies for multi-axis manufacturing. Furthermore, extending this approach to 3D models will require incorporating anisotropic material characteristics arising from scan paths within each layer. Additionally, future studies will investigate multi-physics effects, such as thermal expansion and stress, leveraging the flexibility of OpenMDAO to enhance the applicability of the framework to practical additive manufacturing processes.
\section*{CRediT authorship contribution statement}
\textbf{Seungheon Shin}\: Conceptualization, Methodology, Software, Investigation, Validation, Data curation, Formal analysis, Visualization, Writing - original draft. \textbf{Byeonghyeon Goh}\: Formal analysis, Writing - review \& editing. \textbf{Youngtaek Oh}\: Formal analysis, Writing - review \& editing. \textbf{Hayoung Chung}\: Conceptualization, Writing - original draft, Writing - review \& editing, Resources, Supervision, Project administration, Funding acquisition.
\section*{Declaration of competing interest}
 The authors declare that they have no known competing financial interests or personal relationships that could have appeared to influence the work reported in this paper.
\section*{Acknowledgment}
This research was supported by the National R\&D Program through the National Research Foundation of Korea (NRF), funded by the Ministry of Science and ICT (No. NRF-2020R1C1C1005741). It was also supported by the Korea Institute of Energy Technology Evaluation and Planning (KETEP) by the Ministry of Trade, Industry \& Energy (MOTIE) of the Republic of Korea (No. RS-2023-00240918). Additional support was provided by the National Research Foundation of Korea (NRF), funded by the Korean government (MSIT) (No. RS-2023-00257666).
\appendix
\section*{Appendix A. Sensitivity analysis}
The sensitivities of the overhang angle constraint and material orientation in OpenMDAO are analyzed.
\subsection*{A.1. Sensitivity of overhang angle constraint component}
\renewcommand{\theequation}{A.\arabic{equation}}
\setcounter{equation}{0}
The input and output variables for the overhang angle constraint component are as follows:
\begin{equation}
\begin{aligned}
    &\text{Input:} \ \bm{\rho}, \bm{t}, \bm{\theta} \\ 
    &\text{Output:} \ P_{\bar{\alpha}}
\end{aligned}
\end{equation}
The sensitivity of the overhang angle constraint with respect to the density input variable is derived as below:
\begin{equation}
\begin{aligned}
    \frac{\partial P_{\bar{\alpha}}}{\partial \rho_e} = 
    {
    \sum_{j=1}^N 
    \left(
    \sum_{i \in \mathcal{N}_e}
    \left(
    \frac{\partial H(\xi_i)}{\partial \rho_e} \, \bm{b}^{\{j\}} \cdot \nabla \rho_i^{\{j\}} 
    + H(\xi_i) \, \bm{b}^{\{j\}} \cdot \frac{\partial \nabla \rho_i^{\{j\}}}{\partial \rho_e} 
    \right)\Delta \rho_i^{\{j\}}
    + H(\xi_e) \, \bm{b}^{\{j\}} \cdot \nabla \rho_e^{\{j\}} 
    \frac{\partial \Delta \rho_e^{\{j\}}}{\partial \rho_e}
    \right)
    }
\end{aligned}
\end{equation}
Here, $\Delta \rho^{\{j\}} = \rho^{\{j\}} - \rho^{\{j-1\}}$ represents the density difference between stage $j$ and $j-1$ for each considered element, while $\partial H(\xi_i) / \partial \rho_e$ is given by:
\begin{equation}
\begin{aligned}
    \frac{\partial H(\xi_i)}{\partial \rho_e} 
    &= \frac{\partial H(\xi_i)}{\partial \xi_i} \cdot \frac{\partial \xi_i}{\partial \rho_e} &= \left( \frac{\beta e^{-\beta \xi_i}}{(1 + e^{-\beta \xi_i})^2}  \right) \cdot \frac{\partial \xi_i}{\partial \rho_e}
\end{aligned}
\end{equation}
The term $\partial \xi_i / \partial \rho_e$ is further expanded to reflect its dependence on the density gradient:
\begin{equation}
\begin{aligned}
    \frac{\partial \xi_i}{\partial \rho_e} = 
    \bm{b}^{\{j\}} \cdot 
    \left(
    \frac{
    \frac{\partial \nabla \rho_i^{\{j\}}}{\partial \rho_e} \|\nabla \rho_i^{\{j\}}\|_2 
    - \nabla \rho_i^{\{j\}} \frac{\partial \|\nabla \rho_i^{\{j\}}\|_2}{\partial \rho_e}
    }{
    (\|\nabla \rho_i^{\{j\}}\|_2)^2
    }
    \right)
\end{aligned}
\end{equation}
The derivative of $\partial \nabla \rho_i^{\{j\}} / \partial \rho_e$ and $\partial \|\nabla \rho_i^{\{j\}}\|_2/\partial \rho_e$, which represent the differentiation of the density gradient, are expressed as follows:
\begin{equation}
\begin{aligned}
    \frac{\partial \nabla \rho_i^{\{j\}}}{\partial \rho_e} = \bar{t}_e^{\{j\}} \left(h_1(p, q), h_2(p, q)\right)
\end{aligned}
\end{equation}
and
\begin{equation}
\begin{aligned}
    \frac{\partial \|\nabla \rho_i^{\{j\}}\|_2}{\partial \rho_e} = 
    \frac{\bar{t}_e^{\{j\}} \left(G_1 \cdot h_1(p, q) + G_2 \cdot h_2(p, q)\right)}{\|\nabla \rho_i^{\{j\}}\|_2}
\end{aligned}
\end{equation}
The variables $p$ and $q$ are defined as the relative positions of neighboring elements concerning the element.
The last term $\partial \Delta \rho_e^{\{j\}}/\partial \rho_e$, which involves the product of projected time and density, is expressed as:
\begin{equation}
\begin{aligned}
    \frac{\partial \Delta \rho_e^{\{j\}}}{\partial \rho_e}=
    \frac{\partial (\rho_e^{\{j\}} - \rho_e^{\{j-1\}})}{\partial \rho_e} = \bar{t}_e^{\{j\}} - \bar{t}_e^{\{j-1\}}
\end{aligned}
\end{equation}
Next, the sensitivity of the overhang angle constraint with respect to the time input variable is derived as follows:
\begin{equation}
\begin{aligned}
    \frac{\partial P_{\bar{\alpha}}}{\partial t_e} = 
    {\sum_{j=1}^N
    \left(
    \sum_{i \in \mathcal{N}_e}
    \left(
    \frac{\partial H(\xi_i)}{\partial t_e} \, \bm{b}^{\{j\}} \cdot \nabla \rho_i^{\{j\}}  
    + H(\xi_i) \, \bm{b}^{\{j\}} \cdot \frac{\partial \nabla \rho_i^{\{j\}}}{\partial t_e} 
    \right)\Delta \rho_i^{\{j\}}
    + H(\xi_e) \, \bm{b}^{\{j\}} \cdot \nabla \rho_e^{\{j\}} 
    \frac{\partial \Delta \rho_e^{\{j\}}}{\partial t_e}
    \right)},
\end{aligned}
\end{equation}
where, $\partial H(\xi_i)/\partial t_e$ is:
\begin{equation}
\begin{aligned}
    \frac{\partial H(\xi_i)}{\partial t_e} = 
    \left( \frac{\beta e^{-\beta \xi_i}}{(1 + e^{-\beta \xi_i})^2}  \right) 
    \cdot \frac{\partial \xi_i}{\partial t_e}
\end{aligned}
\end{equation}
The term $\partial \xi_i/\partial t_e$, which depends on time, is expressed as:
\begin{equation}
\begin{aligned}
\frac{\partial \xi_i}{\partial t_e} = 
\bm{b}^{\{j\}} \cdot 
\left(
\frac{
\frac{\partial \nabla \rho_i^{\{j\}}}{\partial t_e} \|\nabla \rho_i^{\{j\}}\|_2 
- \nabla \rho_i^{\{j\}} \frac{\partial \|\nabla \rho_i^{\{j\}}\|_2}{\partial t_e}
}{
(\|\nabla \rho_i^{\{j\}}\|_2)^2
}
\right)
\end{aligned}
\end{equation}
The derivatives $\partial \nabla \rho_i^{\{j\}}/\partial t_e$ and $\partial \|\nabla \rho_i^{\{j\}}\|_2/\partial t_e$ are computed based on the relative positions of neighboring elements, as shown below:
\begin{equation}
\begin{aligned}
\frac{\partial \nabla \rho_i^{\{j\}}}{\partial t_e} = 
\rho_e \frac{\partial \bar{t}_e^{\{j\}}}{\partial t_e} \left(h_1(p, q), h_2(p, q)\right)
\end{aligned}
\end{equation}
and
\begin{equation}
\begin{aligned}
    \frac{\partial \|\nabla \rho_i^{\{j\}}\|_2}{\partial t_e} = 
    \rho_e \frac{\partial \bar{t}_e^{\{j\}}}{\partial t_e} 
    \frac{G_1 \cdot h_1(p, q) + G_2 \cdot h_2(p, q)}{\|\nabla \rho_i^{\{j\}}\|_2}
\end{aligned}
\end{equation}
The term $\partial \bar{t}_e^{\{j\}}/\partial t_e$, which represents the derivative of the Heaviside projection, is expanded into:
\begin{equation}
\begin{aligned}
    \frac{\partial \bar{t}_e^{\{j\}}}{\partial t_e} &= 
    \beta_t \left(\frac{\tanh^2\left(\beta_t (\tau_j - t_e)\right) - 1}{\tanh(\beta_t \tau_j) + \tanh\left(\beta_t (1 - \tau_j)\right)}\right)
\end{aligned}
\end{equation}
The derivative with respect $t_e$ to of the product of projected time and density is given by:
\begin{equation}
\begin{aligned}
    \frac{\partial \Delta \rho_e^{\{j\}}}{\partial t_e}=
    \frac{\partial (\rho_e^{\{j\}} - \rho_e^{\{j-1\}})}{\partial t_e} = 
    \rho_e \frac{\partial (\bar{t}_e^{\{j\}} - \bar{t}_e^{\{j-1\}})}{\partial t_e}
\end{aligned}
\end{equation}
Finally, the sensitivity of the overhang angle constraint with respect to the build orientation variable can be expressed as:
\begin{equation}
\begin{aligned}
    \frac{\partial P_{\bar{\alpha}}}{\partial \theta_j} = 
    {\sum_{e \in \mathcal{M}} 
    \left(
    \frac{\partial H(\xi_e)}{\partial \theta_j} \bm{b}^{\{j\}} 
    + H(\xi_e) \frac{\partial \bm{b}^{\{j\}}}{\partial \theta_j}
    \right)
    \cdot \nabla \rho_e^{\{j\}} \Delta \rho_e^{\{j\}}},
\end{aligned}
\end{equation}
where, $\partial H(\xi_e)/\partial \theta_j$ is given by:
\begin{equation}
\begin{aligned}
    \frac{\partial H(\xi_e)}{\partial \theta_j} = 
    \left( \frac{\beta e^{-\beta\xi_e}}{(1 + e^{-\beta \xi_e})^2}  \right) 
    \frac{\partial \xi_e}{\partial \theta_j}
\end{aligned}
\end{equation}
The term $\partial \xi_e/\partial \theta_j$, which is influenced by the build orientation, is expressed as:
\begin{equation}
\begin{aligned}
    \frac{\partial \xi_e}{\partial \theta_j} = 
    \frac{\partial \bm{b}^{\{j\}}}{\partial \theta_j} \cdot 
    \frac{\nabla \rho_e^{\{j\}}}{\|\nabla \rho_e^{\{j\}}\|_2}
\end{aligned}
\end{equation}
The derivative of the vector $\bm{b}^{\{j\}}$ with respect to $\theta_j$ is computed as:
\begin{equation}
\begin{aligned}
    \frac{\partial \bm{b}^{\{j\}}}{\partial \theta_j} = 
    \left(-\sin(\theta_j), \cos(\theta_j)\right)
\end{aligned}
\end{equation}
By combining these equations, the sensitivity of the overhang angle constraint with respect to all relevant input variables is comprehensively evaluated.
\subsection*{A.2. Sensitivity of material orientation component}
The input and output variables for the material orientation component are given below:
\begin{equation}
\begin{aligned}
    &\text{Input:} \ \bm{t}, \bm{\theta} \\
    &\text{Output:} \ \bm{\phi}
\end{aligned}
\end{equation}
The sensitivities of the material orientation with respect to the input variables are given as follows:
\begin{equation}
\begin{aligned}
    \frac{\partial \phi_e}{\partial t_e} = \sum_{j=1}^N \frac{\partial(\bar{t}_e^{\{j\}} - \bar{t}_e^{\{j-1\}})}{\partial t_e} (\theta_j + \frac{\pi}{2})
\end{aligned}
\end{equation}
and
\begin{equation}
\begin{aligned}
    \frac{\partial \phi_e}{\partial \theta_j} = \bar{t}_e^{\{j\}} - \bar{t}_e^{\{j-1\}}
\end{aligned}
\end{equation}
These equations allow for an accurate computation of the sensitivity of the material orientation to relevant input variables.
\section*{Data availability}
Data will be made available on request.

\bibliography{main}

\end{document}